\documentclass[12pt,a4paper]{article}
\usepackage[a4paper, left=1.5cm, top=3.8cm, right=1.5cm, bottom=3.25cm, marginparsep=5pt, marginparwidth=20mm, headheight=22.53pt]{geometry}
\usepackage{authblk}
\usepackage{float}
\usepackage[T1]{fontenc}
\usepackage[utf8]{inputenc}
\usepackage{amsmath,amsfonts,amsthm}
\usepackage{xcolor}
\usepackage{graphicx}
\usepackage{chngcntr}
\usepackage{caption}
\usepackage{subcaption}
\usepackage{amsthm}
\usepackage{mathrsfs}
\usepackage{tikz}
\usetikzlibrary{decorations.pathmorphing,patterns}
\usetikzlibrary{calc,angles,quotes}
\usetikzlibrary{intersections,arrows}
\usepackage{bm}
\usepackage{natbib}
\usepackage[labelformat=simple]{subcaption}
\usepackage{gensymb}

\counterwithout{figure}{section}

\def\vperi{\upsilon}
\def \bx{\textbf{x}}
\def\jacobi{J}
\def\KSu{\bm{\mathsf{u}}}
\def\KSUP{\bm{\mathsf{U}}}
\def\eps{\varepsilon}
\def\tildeg{\tilde g}

\def\rsoivar{d}
\def\Rsel{d_*}
\def\Rmax{d_{\rm max}}
\def\rhill{d_{\rm H}}
\def\rlapl{d_{\rm L}}
\def\rcheb{d_{\rm C}}
\def\rsoi{d_{\rm soi}}

\def\radEarth{R_{\oplus}}

\def\smaH{a_{\odot}}
\def\eccH{e_{\odot}}
\def\inclH{i_{\odot}}

\def\smaG{a_{\oplus}}
\def\eccG{e_{\oplus}}
\def\inclG{i_{\oplus}}

\def\asymptfPpcm{\nu_{P,a}}
\def\exitfPpcm{\nu_{P,d}}

\def\smaP{a_q}
\def\eccP{e_q}
\def\inclP{i_q}
\def\omP{\omega_q}
\def\OMP{\Omega_q}

\def\q3bp{q}

\def\qpcm{{q_P}}

\def\eccPpcm{e_P}

\def\fPpcm{\nu_P}
\def\fPpcmmin{\nu_{P,\rm min}}

\def\qhelio{q_{\odot}}

\theoremstyle{remark}

\title{\bf A dynamical definition of the sphere of influence of the Earth}
\author[1]{I. Cavallari}
\author[1]{C. Grassi}
\author[1]{G. F. Gronchi}
\author[1]{G. Ba\`u}
\author[2,3]{G.B. Valsecchi}
\affil[1]{Dipartimento di Matematica, Universit\`a di Pisa}
\affil[2]{Istituto di Astrofisica e Planetologia Spaziali (IAPS-INAF)}
 \affil[3]{Istituto di Fisica Applicata Nello Carrara (IFAC-CNR)}

\begin{document}

\maketitle

\begin{abstract}
The concept of sphere of influence of a planet is useful in both the
context of impact monitoring of asteroids with the Earth and of
the design of interplanetary trajectories for spacecrafts.  After
reviewing the classical results, we propose a new definition for this
sphere that depends on the position and velocity of the small body for
given values of the Jacobi constant $C$.  Here we compare the orbit of
the small body obtained in the framework of the circular restricted
three-body problem, with orbits obtained by patching two-body
solutions.  Our definition is based on an optimisation process,
minimizing a suitable target function with respect to the assumed radius of
the sphere of influence.  For different values of $C$ we represent the
results in the planar case: we show the values of the selected radius
as a function of two angles characterising the orbit.  In this case,
we also produce a database of radii of the sphere of
influence for several initial conditions, allowing an interpolation.
\vskip 1mm
\hskip 1cm{\em On peut donc, dans le calcul des
	perturbations d'une com\`ete qui approche tr\`es-pr\`es d'une
	plan\`ete, supposer \`a la plan\`ete une sph\`ere d'activit\'e dans
	laquelle le mouvement relatif de la com\`ete n'est soumis qu'\`a
	l'attraction de la plan\`ete, et au-del\`a de laquelle le mouvement
	absolu de la com\`ete autour du soleil n'est soumis qu'\`a l'action
	du soleil} (P. S. Laplace, Trait\'e de M\'ecanique C\'eleste)
\end{abstract}

\section{Introduction}
The concept of gravitational sphere of influence of a celestial body
was first introduced by \citet{Laplace1805}
to investigate close encounters of comets with Jupiter and the Earth.
In particular, Laplace applied his theory to the study of Lexell's
comet which, in the second half of the 18-th century, had close
encounters with Jupiter and the Earth, producing important changes in
its orbital elements. \citet{Leverrier1,Leverrier2,Leverrier3} was able to
describe the possible changes in the elements \citep[see][]{Valsecchi_etal_2003,Valsecchi2007}, introducing a first
simple version of the Line of Variation \citep[see][]{Milani2005a,Milani2005b}. 

The gravitational effect of a planet on the motion of a small body
was also studied by \citet{TisserandMC} for the capture of
parabolic comets, and by \citet{Fermi1922} in the context of
hyperbolic encounters.

\citet{Opik1976} introduced a model of close encounters, allowing to
explicitly compute the changes in the orbit of the small body after
the encounter. In his model, the encounters are instantaneous, with
intersecting orbits. This mechanism was later generalised by
\citet{Valsecchi_etal_2003} for instantaneous encounters between
objects on orbits with non-zero minimum distance.

The {\em patched-conic} technique is a further
generalisation that allows the encounter to last for a non-zero time
interval, see \citet{Bateetal}.

Besides the motion of natural bodies, the idea of sphere of influence
is also useful in the study of planetary flybys of artificial bodies,
like a spacecraft. This was first investigated by
\citet{Crocco1956}, who studied the possibility of an Earth-Mars-Venus-Earth trajectory using the patched-conic technique. 

The definitions of radius of sphere of influence that are commonly used for a planet are the
following (see Appendix \ref{app:rsoi}):
\begin{align}
&\rlapl =\rho \Bigl(\frac{m_{2}}{m_{1}}\Bigr)^{2/5}, \qquad
\mbox{Laplace's radius} \label{rlaplace}\\
&\rhill = \rho\Bigl(\frac{m_{2}}{3m_{1}}\Bigr)^{1/3},\qquad
\mbox{Hill's radius} \label{rhill}
\end{align}
where $m_1$ is the mass of the Sun, $m_2$ and $\rho$ are respectively the mass and the heliocentric distance of the planet. 
In the case of the Earth we have
\begin{equation*}
\rlapl \approx 0.006\, \mbox{au}, \qquad \rhill \approx 0.01\, \mbox{au}.
\end{equation*}
Both definitions \eqref{rlaplace}, \eqref{rhill} give a
value of the radius depending only on the mass ratio and on the
distance between the planet and the Sun. The same holds true 
for a less-known definition of sphere of influence given
by \citet{Chebotarev1964}. However, from more recent
studies it results that the classical definitions based on the mass ratio
are not always suitable to be employed for the
patched-conic technique. \citet{Araujo_2008} performed a numerical study
showing that the definition of sphere of influence should also depend
on the initial relative velocity between the planet and the
small body. The authors give an empirical law for the radius
of the sphere, but their definition is not applicable to the Earth. Through a numerical study \citet{Amato2017} show that
the patched-conic approximation is more accurate by employing a
sphere of influence different from classical ones, which however
depends on the specific physical problem. In particular, for the
Sun-Earth problem they found that the most suitable radius lies
between $1.2\rhill$ and $3\rhill$.

This ambiguity on the definition of the sphere of influence,
confirmed by the recent works, is a well-known issue. Laplace himself observed that it is possible to take
larger values of the radius of the sphere with respect to the one
computed from his formula, and still obtain good results:

\vskip 0.3cm \hskip 1cm{\em On peut m\^eme beaucoup augmenter le
	rayon de cette sph\`ere, sans qu'il en r\'esulte d'erreur sensible.}
\citep[][Chap. 2, Book IX]{Laplace1805}
\vskip 0.3cm

The purpose of this paper is to unravel the ambiguity and determine
the most suitable sphere of influence for the Earth-Sun problem for
the patched-conic method, such that the main features of a close
encounter and of the post-encounter trajectory are well reproduced. In light of the results cited above, our definition
takes into account also the position and velocity of the small
body.

The paper is organized as follows. After setting the notation and
recalling some basic properties of the restricted three-body problem in
Section~\ref{sec:CR3BP},  we describe some features of
close encounters in Section~\ref{sec:closenc}. We discuss the patched-conic method in Section~\ref{sec:patch}. In
Section~\ref{sec:newdef} we introduce the procedure used to define our
sphere of influence focusing on the planar case. Some conclusions and
a comparison with the classical definitions $\rlapl$ and $\rhill$ are
presented in Section~\ref{sec:results}.  Additional material is
included in the Appendix, where we also discuss the 3-dimensional
case.

\section{Dynamical model} 
\label{sec:CR3BP}
For our purpose, we take the dynamical model of the circular restricted
three-body problem (CR3BP). We consider a small body, modelled as a
massless particle, moving under the gravitational influence of the Sun
and the Earth, also called primaries of the problem, see
\citet{Koon_2011}. The primaries are assumed rotating on circular
orbits around their common centre of mass.

We make the system non-dimensional. Let $m_1$ and
$m_{2}$ be the masses of the Sun and the Earth. The
$m_1 + m_2$ is chosen as unit of mass. In this way, the non-dimensional masses
of the Sun and the Earth are respectively $1-\mu$ and $\mu$, where
\begin{equation*}
\mu = \frac{m_{2}}{m_{1} + m_{2}}.
\end{equation*}
The distance between the Earth and the Sun is taken as unit of
length. Finally, the unit of time is set so that the orbital period of
the Earth and the Sun around their centre of mass is equal to
$2\pi$. Thus, the universal constant of gravitation becomes equal to
$1$.

Let us consider a barycentric synodic reference frame $Oxyz$ rotating with
angular velocity equal to the mean motion of the Earth and the Sun,
i.e. equal to $1$. The $(x,y)$-plane corresponds to the plane of
motion of the primaries. The $x$-axis lies on the line passing through
the primaries and it points towards the Earth. The $z$-axis has the
direction of the angular momentum of the primaries. In this reference
frame the coordinates of the Sun and the Earth are $(-\mu,0,0)$ and
$(1-\mu,0,0)$, respectively.

The Hamiltonian function of the CR3BP is
\begin{equation}
H_R = \frac{p_{x}^2+p_{y}^2+p_{z}^2}{2} + p_{x}y-p_{y}x
-\frac{1-\mu}{r}-\frac{\mu}{d},
\label{3bpHam}
\end{equation}
where
\begin{equation}
r = \sqrt{(x+\mu)^2+y^2+z^2}
\label{rdef}
\end{equation}
is the distance between the particle and the centre of the Sun, and
\begin{equation}
d = \sqrt{(x-1+\mu)^2+y^2+z^2}
\label{ddef}
\end{equation}
is the distance between the particle and the centre of the Earth. In
the Hamiltonian function, $p_x$, $p_y$, $p_z$ are the momenta
conjugated to $x$, $y$, $z$, and fulfil
\begin{equation*}
p_x = \dot{x} - y, \quad p_y = \dot{y} + x, \quad p_z = \dot{z}.
\end{equation*}
$H_R$ is an integral of motion. In particular,
\begin{equation}
\jacobi=-2H_R =
2\frac{(1-\mu)}{r}+2\frac{\mu}{d}+x^2+y^2-(\dot{x}^2+\dot{y}^2+\dot{z}^2)
\label{Cdef}
\end{equation}
is called Jacobi integral.  This quantity can be expressed in two
alternative forms, employing the heliocentric and geocentric Keplerian
osculating elements of the small body:
\begin{equation}
\jacobi=\frac{1-\mu}{\smaH}+2\sqrt{(1-\mu)\smaH(1-\eccH^2)}\cos
\inclH+2\mu\Bigl(\frac{1}{d}-r\cos\alpha\Bigr)+\mu^2\, ,
\label{Chelio}
\end{equation}
where $\smaH, \eccH, \inclH$ are the heliocentric semi-major axis,
eccentricity and inclination and $\alpha$
is the angle between the small body and the planet seen from the Sun;

\begin{equation}
\jacobi= -\frac{\mu}{\smaG} + 2\sqrt{\mu \smaG(\eccG^2-1)}\cos \inclG +
2(1-\mu)\left(\frac{1}{r} - d\cos\varphi\right) + (1-\mu)^2,
\label{Cplaneto}
\end{equation}
where $\smaG,\eccG,\inclG$ are the geocentric semi-major
axis, eccentricity and inclination, and $\varphi$ is the angle between
the small body and the Sun seen from the planet.  In Appendix
\ref{app:jacobi} we show the computations leading to the three
expressions for $J$ given in \eqref{Cdef}, \eqref{Chelio} and
\eqref{Cplaneto}.

\begin{figure}
	\centering
	\includegraphics{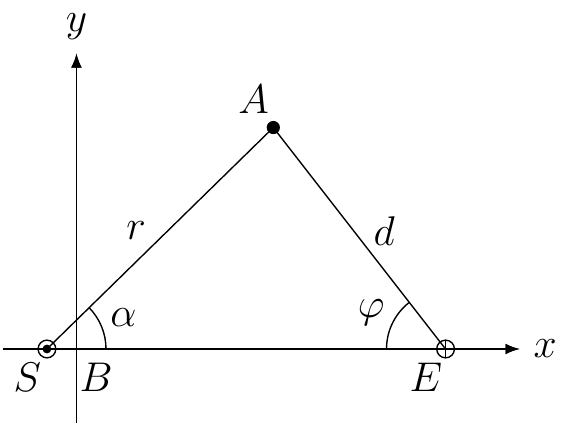}
	\caption{Description of the positions of the Sun ($S$), the Earth
		($E$) and the small body ($A$) in the barycentric synodic
		reference frame.}
\end{figure}

\section{Close encounters}
\label{sec:closenc}
The orbit of the small body can be approximated as a Keplerian
planetocentric hyperbola when it is close to the Earth and it is not
captured by the gravitational attraction of the planet. Indeed, in the
neighbourhood of the planet, the gravitational influence of the Sun
becomes negligible. With this hypothesis, in the following we discuss
some general properties of close encounters.

\subsection{Types of encounters}
\label{sec:enctype}

\begin{figure}
	\centering
	\begin{tikzpicture}
	\coordinate (A) at ($(-4,0)$);
	\coordinate (B) at ($(0,0)$);
	
	\coordinate (C) at ($({-5*cos(30)},{5*sin(30)})$);
	\coordinate (O) at ($({-2.5*cos(30)},{2.5*sin(30)})$);
	
	\coordinate (D) at ($({-5*cos(30)},{-5*sin(30)})$);
	\coordinate (Q) at ($({(-5*cos(30))/2},{(-5*sin(30))/2})$);
	
	\coordinate (E) at ($({-4*sin(60)*cos(30)},{4*sin(60)*sin(30)})$);
	\coordinate (S) at ($({-4*sin(60)*cos(15)},{4*sin(60)*sin(15)})$);
	
	\coordinate (F) at ($({-4*sin(60)*cos(30)},{-4*sin(60)*sin(30)})$);
	\coordinate (T) at ($({-4*sin(60)*cos(15)},{-4*sin(60)*sin(15)})$);
	
	\draw [-latex] [line width=0.25mm] (B) -- (C);
	\draw [-latex] [line width=0.25mm] (D) -- (B);
	\draw [dashed] (A) -- (B);
	\draw [] (A) -- (E);
	\draw [] (A) -- (F);
	
	\draw (O) node [above=0.1cm] {$U'$};
	\draw (Q) node [below=0.1cm] {$U$};
	\draw (S) node [left=0.2cm] {$b$};
	\draw (T) node [left=0.2cm] {$b'$};
	
	\pic [draw, "$\gamma$", angle eccentricity=1.4, angle radius = 0.6cm] {angle = F--A--B};
	\pic [draw, angle eccentricity=1.4, angle radius = 0.6cm] {angle = B--A--E};
	\end{tikzpicture}
	\caption{The incoming asymptotic velocity $U$ is
		deflected by an angle $\gamma$ into the outgoing asymptotic
		velocity $U'$.}
	\label{fig:deflectionAngle}
\end{figure}
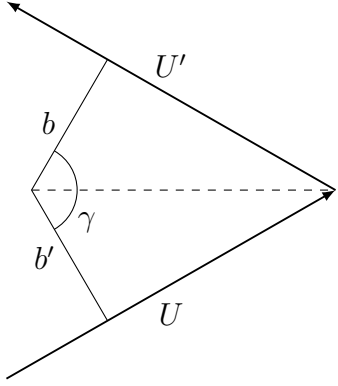

We classify close encounters on the basis of two quantities in the framework of the CR3BP: 
\begin{enumerate}
	\item the minimum distance $\q3bp$ between the planet and the
	small body;
	\item the magnitude of the relative velocity $\vperi$ of the
	small body at the time $t_{\q3bp}$ corresponding to the
	minimum distance $q$.
\end{enumerate}
In particular, we distinguish between {\em deep} and {\em shallow}
encounters depending on the values of $\q3bp$, and between {\em fast}
and {\em slow} encounters depending on the value of $\vperi$.

\begin{figure}
	\centering
	\includegraphics[width=0.7\textwidth]{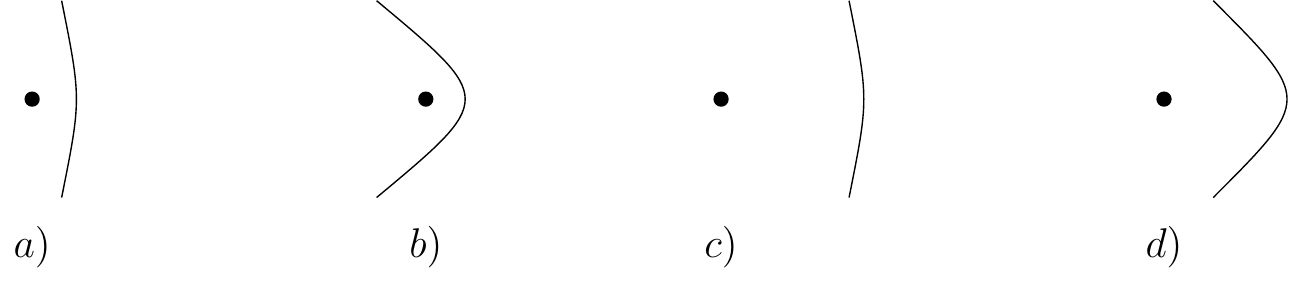}
	\caption{The four possible cases: $a)$ deep straight, $b)$
		deep curved, $c)$ shallow straight, $d)$ shallow
		curved. Here the small disk represents the Earth.}
	\label{fig:enctype}
\end{figure}

Let us consider the osculating Keplerian orbit at $t=t_{\q3bp}$. The
following relations hold between $\q3bp, \vperi$ and $U, b$, where $U$
is the magnitude of the asymptotic velocity of the hyperbolic orbit and $b$ is its impact
parameter \citep[see][]{Valsecchi_etal_2003}:
\begin{equation}
U^2 = \frac{\mu}{\q3bp}\left(\frac{\q3bp\vperi^2}{\mu} - 2\right),
\qquad b = \q3bp\frac{\sqrt{(\frac{\q3bp\vperi^2}{\mu} -
		1)^2-1}}{\frac{\q3bp\vperi^2}{\mu} - 2}.
\label{Ubdminvperi}
\end{equation}
From relations
\begin{equation*}
U^2 = \vperi^2 - \frac{2\mu}{\q3bp}, \qquad \q3bp = \smaP(\eccP-1),
\qquad \smaP = \frac{\mu}{U^2},
\end{equation*}
we obtain
\begin{equation}
\eccP = \frac{\q3bp\vperi^2}{\mu} - 1,
\label{evperi}
\end{equation}
so that we can write \eqref{Ubdminvperi} as
\begin{equation}
U^2 = \frac{\mu}{\q3bp}(\eccP-1),\qquad
b = \q3bp\sqrt{\frac{\eccP+1}{\eccP-1}}.
\label{Ubdperie}
\end{equation}
Here, $\smaP=\smaG(t_{\q3bp})$, $\eccP=\eccG(t_{\q3bp})$.  Relation
\eqref{evperi} implies that for a fixed value of $\q3bp$, the
eccentricity $\eccP$ is increasing if $\vperi$ is increasing. Let us
also remark that from relation $\sin(\gamma/2) = {1}/{\eccP}$,
where $\gamma$ is the deflection angle (see Fig.
\ref{fig:deflectionAngle}), we obtain that the greater the
eccentricity $\eccP$ is, the less the trajectory is curved.  As a
consequence, for a fixed value of $\q3bp$, faster and slower
encounters correspond to more straight and more curved hyperbolic
trajectories, respectively. In Figure \ref{fig:enctype} we draw a
sketch of the four possible cases for the trajectories. We will see in
Section~\ref{sec:ep} that, depending on the value of the Jacobi
constant $C$, some of these combinations will not be possible.

\subsubsection{Estimates for $\eccP$}
\label{sec:ep}
We derive optimal estimates for the values of the geocentric
eccentricity $\eccP$ at $t_{\q3bp}$ depending on a chosen value of the
Jacobi constant $C$.

\begin{figure*}
	\begin{subfigure}{.48\textwidth}
		\centering
		\includegraphics[width=\textwidth]{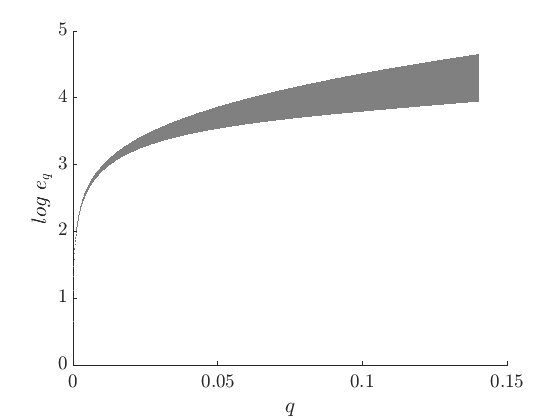}
	\end{subfigure}%
	\begin{subfigure}{.48\textwidth}
		\centering
		\includegraphics[width=\textwidth]{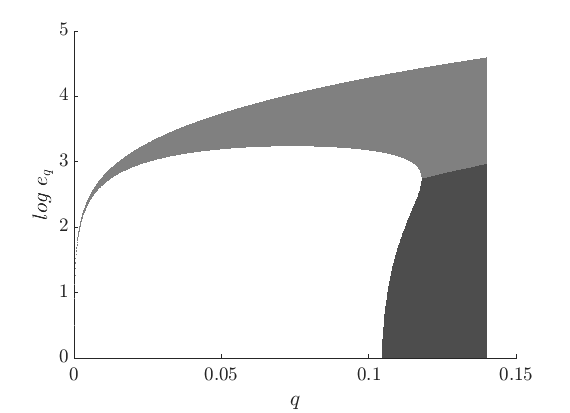}
	\end{subfigure}
	\begin{subfigure}{.48\textwidth}
		\centering
		\includegraphics[width=\textwidth]{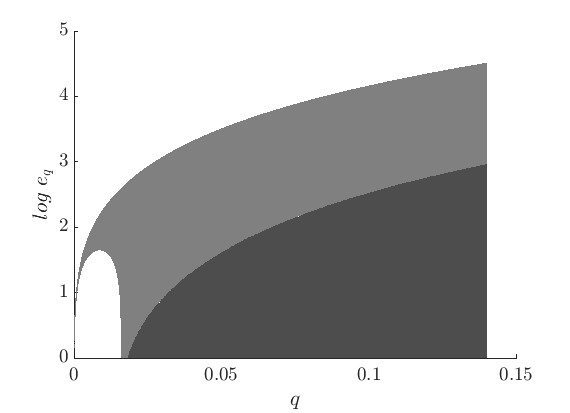}
	\end{subfigure}%
	\begin{subfigure}{.48\textwidth}
		\centering
		\includegraphics[width=\textwidth]{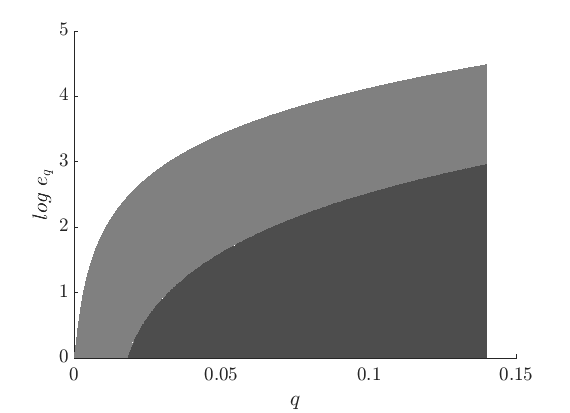}
	\end{subfigure}
	\caption{Possible values of $\eccP$ (logarithmic scale) for
		$C=2.75$ (top left), $C=2.85$ (top right), $C=2.97$ (bottom
		left) and $C=2.999$ (bottom right). The allowed region is coloured
		in grey. In the light grey region $\eccP={\eccP}_{_1}$, while in the dark
		grey region $\eccP={\eccP}_{_2}$.}
	\label{eccInterval}
\end{figure*}

Let $\inclP,\OMP,\omP$ 
be the osculating inclination, longitude of the node and argument of perigee at $t_{\q3bp}$.  The distance from the Sun is
\begin{equation*}
r = \sqrt{1+\q3bp^2-2\q3bp\cos\varphi}\, ,
\end{equation*}
with 
\begin{equation*}
\cos\varphi = -\cos(\omP)\cos(\OMP-\lambda) +
\cos(\inclP)\sin(\omP)\sin(\OMP-\lambda),
\end{equation*}
where $\lambda$ is the longitude of the Earth.
At the minimum geocentric distance along the 3-body orbit,
equation \eqref{Cplaneto} can be written as
\begin{equation*}
s(C,\varphi,\q3bp) +\frac{\mu
	\eccP}{\q3bp}-2\sqrt{\mu\q3bp(\eccP+1)}\cos \inclP=0,
\label{eccEq}
\end{equation*}
with
\begin{equation}
s(C,\varphi,\q3bp)  =   C -
2(1-\mu)\left(\frac{1}{\sqrt{1+\q3bp^2-2\q3bp \cos\varphi}}-\q3bp
\cos\varphi\right)\quad -(1-\mu)^2-\frac{\mu}{\q3bp}.
\end{equation}
From \eqref{eccEq} we get that for $\inclP=\frac{\pi}{2}$, given $C,
\varphi$ and $\q3bp$, the possible values of $\eccP$ are simply
\begin{equation*}
\eccP= -\frac{\q3bp}{\mu}s.
\end{equation*}
Instead, for $\inclP\ne\frac{\pi}{2}$, we get the possible values of
$\eccP$ by solving
\begin{equation*}
2\sqrt{\mu\q3bp(\eccP+1)}\cos \inclP = \frac{\mu}{\q3bp}\eccP +
s.
\end{equation*}
Then, we have $\eccP = {\eccP}_{_{1,2}}$ with
\begin{equation}
\begin{cases}
{\eccP}_{_{1}} = \frac{\q3bp}{\mu}(2\q3bp^2\cos^2\inclP-s)
+
2\frac{\q3bp}{\mu}\sqrt{\q3bp\cos^2\inclP\left(\q3bp^3\cos^2\inclP-\q3bp
	s+\mu\right)}, \\[5pt]  {\eccP}_{_{2}} =
\frac{\q3bp}{\mu}(2\q3bp^2\cos^2\inclP-s) -
2\frac{\q3bp}{\mu}\sqrt{\q3bp\cos^2\inclP\left(\q3bp^3\cos^2\inclP-\q3bp
	s+\mu\right)}.
\end{cases} 
\label{eccVal}
\end{equation}
These are acceptable, non-spurious solutions if $\eccP\in\mathbb{R}$ and
if
\begin{equation*}
\begin{cases}
\frac{\mu}{\q3bp}\eccP+s >0 & \text{for}\; 0\le \inclP<\pi/2,\\[5pt]
\frac{\mu}{\q3bp}\eccP+s<0 & \text{for}\; \pi/2 < \inclP\le\pi.
\end{cases}
\end{equation*}
Moreover, since we are looking for hyperbolic orbits, we ask that the
condition $\eccP>1$ is also satisfied.  The possible values of $\eccP$
as a function of the minimum distance $\q3bp$ are shown in Fig.
\ref{eccInterval} for $\inclP=0$. Here, we can see different
possibilities depending on $C$. For
$C=2.75$, ${\eccP}_{_{1}}$ is the only acceptable solution for all
values of $\q3bp$ and the minimum and maximum values of $\eccP$
increase as $\q3bp$ increases. For larger values of $C$, the solutions
${\eccP}_{_{2}}$ can become acceptable starting from a certain value
of $\q3bp$. In this way, we can have hyperbolic orbits with small
eccentricities for $\q3bp$ greater than some $\bar{\q3bp}$ that
depends on $C$. We also note that $\bar{\q3bp}$ decreases as $C$
increases, so that for values of $C$ very close to 3 there are orbits
with small geocentric eccentricities for all values of the perigee
distance $\q3bp$.

\subsection{The Tisserand parameter}
\label{sec:tiss}

\begin{figure}
	\centering
	\includegraphics[width=.48\textwidth]{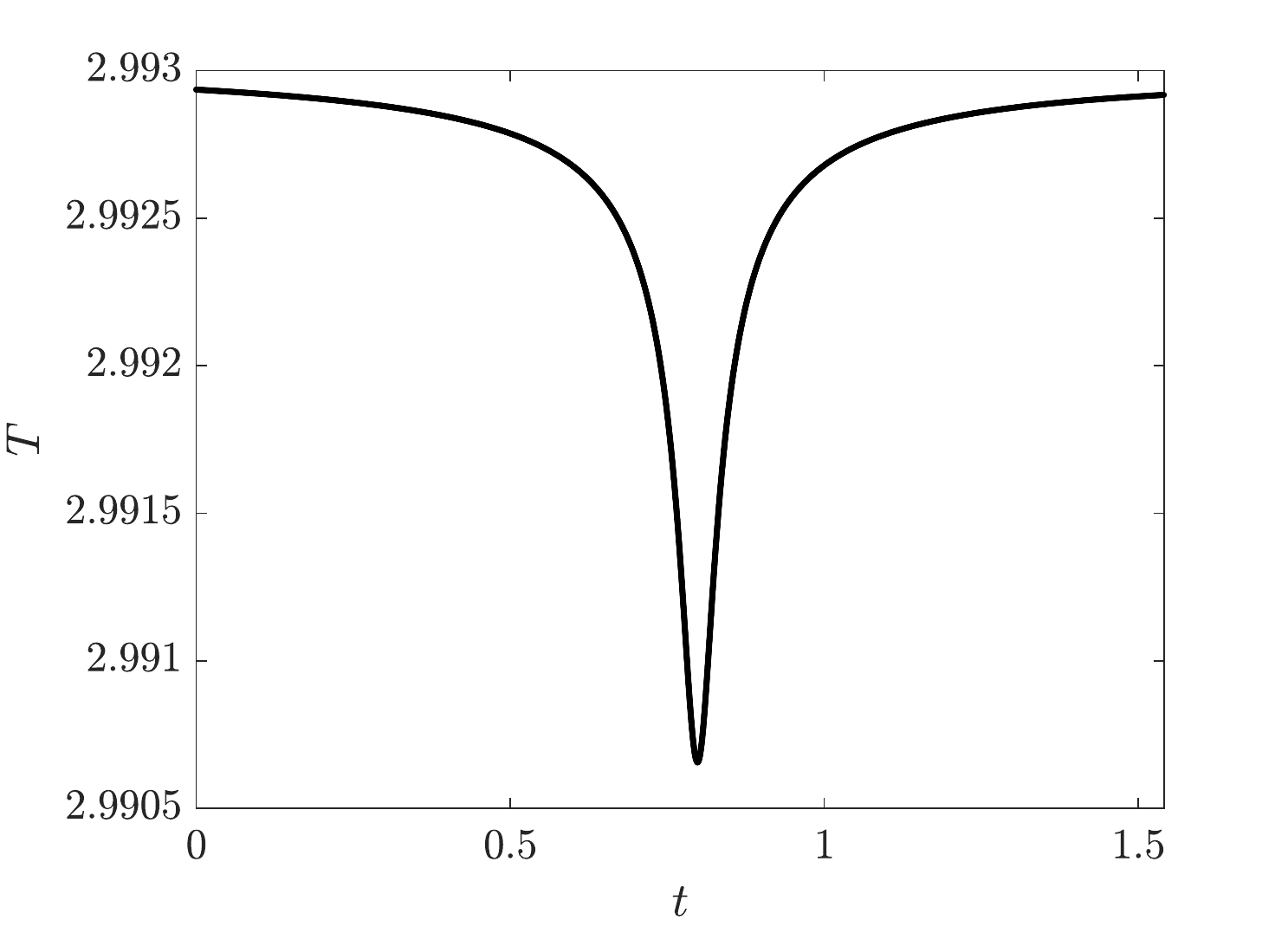}
	\caption{Variation of the value of the Tisserand parameter during a
		close encounter with a planet.}
	\label{Tisserand}
\end{figure}

It is well known that there exists a quantity, called Tisserand
parameter, which takes almost the same value before and after a close
encounter with a planet \citep{Tisserand1889}. Indeed, it remains
almost constant when the small body is far from the planet, it
decreases as the small body approaches the planet and then stabilises
on a value that is very close to the pre-encounter one. The typical
behaviour of the Tisserand parameter during a close encounter is shown
in Fig. \ref{Tisserand}.

In terms of the orbital elements, the Tisserand parameter, denoted by
$T$, is written as
\begin{equation}
T=\frac{1-\mu}{\smaH}+2\sqrt{(1-\mu)\smaH(1-\eccH^2)}\cos \inclH.
\label{TissDef}
\end{equation}

In this section we derive an upper bound for the time-derivative of
$T$. Our aim is to find a distance $d_C$ from the
Earth, depending on the Jacobi constant, where the effect of the planetary
gravitational attraction is negligible. For the optimisation process described
in Section~\ref{sec:newdef} we will consider initial conditions that
depend on $d_C$.
From \eqref{Chelio} and \eqref{TissDef} we get
\begin{equation*}
T=\jacobi-2\mu\left(\frac{1}{d} - r\cos\alpha\right) - \mu^2 = \jacobi
- 2\mu\left(\frac{1}{d} - x\right) + \mu^2\, .
\end{equation*}
In the synodic barycentric reference frame, the position of the small
body can be written in terms of the spherical angles $\eta$ and $\chi$
as
\begin{equation*}
\begin{aligned}
x &= d\cos\eta\cos\chi+1-\mu,\\
y &= d\sin\eta\cos\chi,\\
z &= d\sin\chi,
\end{aligned}
\end{equation*}
where $d$ is the geocentric distance.  From \eqref{Cdef} we see
that the norm $v=\sqrt{\dot x^2+\dot y^2+\dot z^2}$ of the velocity is
given once the value of the Jacobi integral $\jacobi$
and the position of the small body are fixed. 
Then, the velocity
vector of the small body can also be described by two angles
$\theta,\zeta$, so that
\begin{equation*}
\begin{aligned}
\dot x &= v\cos\theta\cos\zeta,\\
\dot y &= v\sin\theta\cos\zeta,\\
\dot z &= v\sin\zeta.
\end{aligned}
\end{equation*}
The time derivative of $T$ is 
\begin{equation*}
\begin{aligned}
\frac{\mathrm{d}T}{\mathrm{d}t}&=2\mu\left(\frac{(x-1+\mu)\dot x+y\dot y+z\dot
	z}{d^3}+\dot x\right)\\ & = 2\mu
v\left(\frac{\cos\chi\cos\zeta\cos(\eta-\theta) +
	\sin\chi\sin\zeta}{d^2} + \cos\theta\cos\zeta\right).
\end{aligned}
\end{equation*}
Since $v>0$, we have
\begin{equation*}
\left\vert\frac{\mathrm{d}T}{\mathrm{d}t}\right\vert = 2\mu
v\left\vert\frac{\cos\chi\cos\zeta\cos(\eta-\theta) +
	\sin\chi\sin\zeta}{d^2} + \cos\theta\cos\zeta\right\vert.
\end{equation*}
Thus, 
\begin{equation*}
\left\vert\frac{\mathrm{d}T}{\mathrm{d}t}\right\vert  \le  2\mu
v\Bigg(\frac{\left\vert\cos\chi\cos\zeta\cos(\eta-\theta) +
	\sin\chi\sin\zeta\right\vert}{d^2}  \quad + \left\vert\cos\theta\cos\zeta\right\vert\Bigg) \le 2\mu v \Bigg(\frac{1}{d^2}+1\Bigg).
\end{equation*}
Let us call $C$ the constant value assumed by the Jacobi integral
$\jacobi$.  Knowing that
\begin{gather}
r^2=(x+\mu)^2+y^2+z^2=1+d^2+2d\cos\eta\cos\chi ,\\
x^2+y^2=d^2\cos^2\chi+2(1-\mu)d\cos\eta\cos\chi+(1-\mu)^2,
\label{rxy_fund}
\end{gather}
from \eqref{Cdef} we obtain
\begin{equation*}
v^2=2\frac{1-\mu}{r}+2\frac{\mu}{d}+x^2+y^2-C
= g(d,\eta,\chi)-C.
\end{equation*}
with 
\begin{equation*}
g(d,\eta,\chi)  =  2\frac{1-\mu}{\sqrt{1+d^2+2d\cos\eta\cos\chi}} 
+2\frac{\mu}{d}+d^2\cos^2\chi\quad +2(1-\mu)d\cos\eta\cos\chi+(1-\mu)^2.
\end{equation*}
It holds $g(d,\eta,\chi) \le \bar{g}(d)$, where
\begin{equation*}
\bar{g}(d) = \frac{2(1-\mu)}{1-d}+2\frac{\mu}{d}+d^2+2(1-\mu)d+(1-\mu)^2.
\end{equation*}
We have $\bar{g}(d)-C>0$ for every value of $C\le 3$, considering that $d\in[\radEarth,1)$, with $\radEarth$ the Earth radius.
\begin{figure}
	\centering
	\includegraphics[width=.48\textwidth]{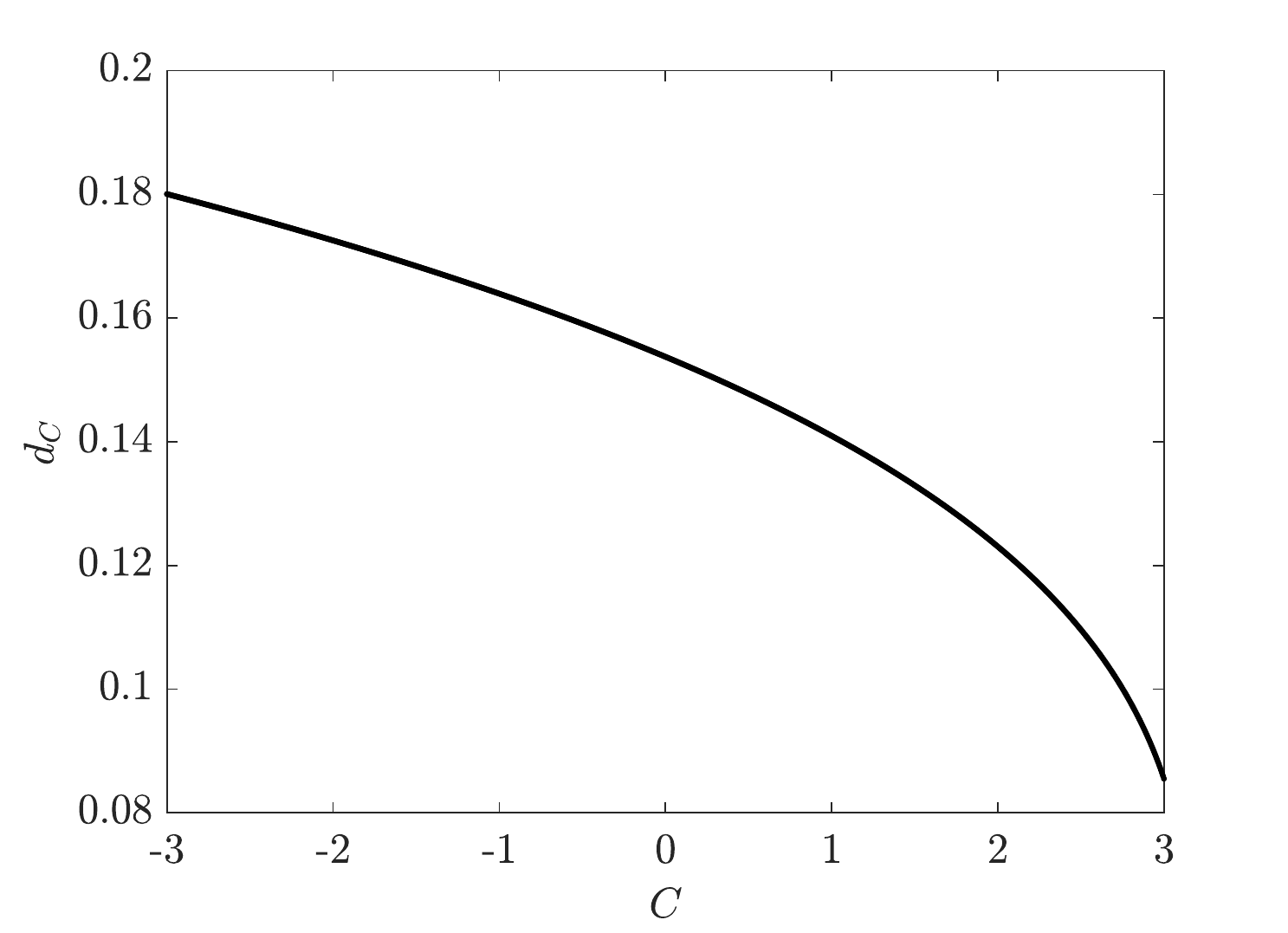}
	\caption{Values of $d_C$ as a function of $C$.}
	\label{fig:dref}
\end{figure}
We can now look for values of the geocentric distance $d$ at which
the Tisserand parameter remains approximately constant, by solving
\begin{equation}
2\mu\sqrt{\bar{g}(d)-C}\left(\frac{1}{d^2}+1\right)
\le\epsilon,
\label{dTdtMinEps}
\end{equation}
with $\epsilon$ some small quantity. This bounding condition
corresponds to the inequality
\begin{equation}
h(C,d)\le0
\label{ineq_dist}
\end{equation}
where
\begin{equation}
\begin{aligned}
h(C,d)=&-d^8+(2\mu-1)d^7+(C+\sigma-\mu^2-1)d^6\\
& +(1-2\mu+\mu^2-C-\sigma)d^5
+(2\mu+1-2\mu^2+2C)d^4\\
& +(5-10\mu+2\mu^2-2C)d^3
+(4\mu+1-\mu^2+C)d^2\\
&+(3-6\mu+\mu^2-C)d+2\mu,
\end{aligned}
\label{hdef}
\end{equation}
with $\sigma={\epsilon^2}/{4\mu^2}$.  
Let us consider $C\in[-3,3]$ and set  $\epsilon=5\cdot10^{-4}$. 
From \eqref{hdef} we have
\begin{equation*}
h(C,0)=2\mu>0,\qquad\lim_{d\to+\infty}h(C,d)=-\infty
\end{equation*}
therefore for any value of $C$ there exists at least one positive
solution of $h(C,d)=0$. Let $d_1(C)<d_2(C)<\dots$ be the positive solutions of
$h(C,d)=0$ and ${\tilde d}_j=d_j(-3)$. From the implicit function
theorem we know that
\begin{equation*}
d_j'(C) = -\frac{\frac{\partial h}{\partial
		C}(C,d_j(C))}{{\frac{\partial h}{\partial d}(C,d_j(C))}}.
\end{equation*}
We have
\begin{equation*}
\frac{\partial h}{\partial C}(C,d(C))=d(d-1)(d^2+1)^2,
\end{equation*}
which is negative for $0<d<1$. Moreover,
\begin{equation*}
\frac{\partial h}{\partial d}(C,d_1(C))<0,\qquad
\frac{\partial h}{\partial d}(C,d_2(C))>0.
\end{equation*}
From a numerical evaluation we get $h(-3,0.5)<0$, therefore it must be ${\tilde d}_1<0.5$.  We then have
$d_1'(C)<0$. Furthermore, $d_1(3)>\radEarth$, so that $d_1(C)>\radEarth$
for all $C\in[-3,3]$. Since $h(C,1)=8(1-\mu)>0$, we
get that there exists $d_2(C)<1$ for all $C\in[-3,3]$. We then also
have $d_2'(C)>0$. Therefore, we conclude that
$\radEarth<d_1(C)\le\tilde d_1$ and $\tilde{d}_2\le d_2(C)<1$ for
every $C\in[-3,3]$. We take $d_C$ as the smallest positive value of $d$ that satisfies inequality \eqref{ineq_dist}, i.e. $d_C=d_1(C)$. In Fig. \ref{fig:dref} we show the evolution of $d_C$ as $C$ varies in the interval $[-3,3]$. 


\section{Patched-conic Method}
\label{sec:patch}
Planetary close encounters can be studied by means of the
patched-conic approximation. The small body is modelled as a massless
particle, travelling along a heliocentric Keplerian elliptic orbit
before and after the encounter. The encounter occurs inside the sphere
of influence of the Earth, where the orbit of the small body is a
geocentric Keplerian hyperbolic orbit.

By adopting the synodic barycentric reference frame introduced in
Section \ref{sec:CR3BP}, the Hamiltonian function of the heliocentric
Kepler problem is
\begin{equation}
H_{\odot} = \frac{p_x^2+p_y^2+p_z^2}{2} + p_{x}y-p_{y}x
-\frac{1-\mu}{r}-\mu x,
\label{Ham_sun}
\end{equation}
with $r$ defined in \eqref{rdef}.
Similarly, the Hamiltonian function of the geocentric Kepler
problem is
\begin{equation}
H_{\oplus}=
\frac{p_x^2+p_y^2+p_z^2}{2}+p_xy-p_yx-\frac{\mu}{d}+(1-\mu)x,
\label{Ham_earth}
\end{equation}
with $d$ defined in \eqref{ddef}.
In the patched-conic method, when the small body is far from the planet
we only consider the gravitational influence of the Sun, so that the
trajectory of the small body is simply described by a Keplerian elliptic
orbit resulting from the dynamics associated to the Hamiltonian
\eqref{Ham_sun}. In case of close encounters with the Earth, if the
geocentric distance of the small body is small enough, the
gravitational attraction of the planet becomes dominant. When this
happens, we instantaneously change the centre of gravity: the small body
is now only affected by the influence of the Earth and its trajectory
is simply given by a Keplerian hyperbolic orbit resulting from the
Hamiltonian dynamics associated to \eqref{Ham_earth}. Then, once the
small body gets far enough from the planet, we switch back again to the
pure Sun-particle two-body problem.

\subsection{Effective deflection}
\label{sec:deflect}
Let ${\cal O}_p$ be the geocentric hyperbolic orbit inside the sphere
of influence.
${\cal O}_p$ will depend on the chosen radius $\rsoi$ of the sphere of
influence. Let us denote by $\hat{\bm v}_d$ the unit vector of
the geocentric velocity at the exit point from the sphere of influence, by
$\hat{\bm v}_a$ the unit vector of the hyperbolic asymptotic
velocity, and by $\Delta\theta$ the angle between $\hat{\bm
	v}_d$ and $\hat{\bm v}_a$.

Here, we derive the relation between $\Delta\theta,\rsoi,\qpcm,\eccPpcm$, where  $\qpcm$ and $\eccPpcm$ are the pericentre distance and the eccentricity of ${\cal O}_p$.

The position and velocity along ${\cal O}_p$ are given by
\begin{equation*}
\begin{cases}
x=d\cos \fPpcm\\
y=d\sin \fPpcm
\end{cases},\qquad
\begin{cases}
x'=d'\cos \fPpcm-d\sin \fPpcm\\
y'=d'\sin \fPpcm+d\cos \fPpcm
\end{cases},
\end{equation*}
with $\fPpcm$ the true anomaly, $d$ the geocentric distance and $d'$
its derivative with respect to $\fPpcm$. In particular,
\begin{equation*}
d=\frac{\qpcm(1+\eccPpcm)}{1+\eccPpcm\cos \fPpcm},\qquad d'=\frac{\qpcm\eccPpcm(1+\eccPpcm)\sin \fPpcm}{(1+\eccPpcm\cos \fPpcm)^2}.
\end{equation*}
The squared magnitude of the velocity along the hyperbolic orbit is
\begin{equation*}
v^2=x'^2+y'^2=\frac{\qpcm^2(1+\eccPpcm)^2}{(1+\eccPpcm\cos
	\fPpcm)^4}\left(1+2\eccPpcm\cos \fPpcm+\eccPpcm^2\right).
\end{equation*}
The asymptotic value $\asymptfPpcm$ of the true anomaly is determined by
\begin{equation*}
\asymptfPpcm=\arccos\left(-\frac{1}{\eccPpcm}\right)
\end{equation*}
so that
\begin{equation*}
\hat{\bm v}_a = \left(\begin{array}{c}
-\frac{1}{\eccPpcm}\\[5pt]\frac{\sqrt{\eccPpcm^2-1}}{\eccPpcm}
\end{array}\right).
\end{equation*}
Instead, the value $\exitfPpcm$ of the true anomaly at the exit of the sphere
of influence is determined by
\begin{equation*}
\exitfPpcm = \arccos\left( \frac{1}{\eccPpcm}\left(\frac{\qpcm(1+\eccPpcm)}{d}-1\right)\right).
\end{equation*}
Therefore, the normalised velocity $\hat{\bm v}_d$ can be written as
\begin{equation*}
\begin{aligned}
\hat{\bm v}_d &= \frac{(1+\eccPpcm\cos \exitfPpcm)^2}{\qpcm(1+\eccPpcm)\sqrt{1+2\eccPpcm\cos \exitfPpcm}}
\left[d'
\begin{pmatrix}
\cos \exitfPpcm\\[5pt]
\sin \exitfPpcm
\end{pmatrix}
+d
\begin{pmatrix}
-\sin \exitfPpcm\\[5pt]
\cos \exitfPpcm
\end{pmatrix}
\right]\\
&=\frac{1}{\sqrt{1+2\eccPpcm\cos \exitfPpcm+\eccPpcm^2}}
\begin{pmatrix}
\sqrt{1-\frac{1}{\eccPpcm^2}\left[\frac{\qpcm(1+\eccPpcm)}{d}-1\right]^2}\\[12pt]
\frac{1}{\eccPpcm}\left[\frac{\qpcm(1+\eccPpcm)}{d}-1\right]+\eccPpcm
\end{pmatrix}.
\end{aligned}
\end{equation*}
We note that $\asymptfPpcm -\frac{\pi}{2} = \frac{\gamma}{2}$, where $\gamma$
is the deflection angle introduced in Section~\ref{sec:enctype}.
The angle between $\hat{\bm v}_d$, $\hat{\bm v}_a$ corresponds
to half the angle of {\em missed} deflection, that we denote by
$\Delta\theta$, so that
\begin{equation*}
\hat{\bm v}_d\cdot\hat{\bm v}_a = \cos\left({\frac{\Delta \theta}{2}}\right).
\end{equation*}

\section{A new definition of sphere of influence}
\label{sec:newdef}
In this section we describe the numerical method used to define a
suitable radius of the sphere of influence, depending on the state
variables of the small body. From now on we restrict our analysis to the
planar problem.

\subsection{Initial Conditions}
\label{sec:ic_planar}

\begin{figure}
	\begin{center}
		\begin{tikzpicture}
		\coordinate (O) at (0,0);
		\coordinate (Od) at (0,-2);
		\coordinate (Ou) at (0,3);
		\coordinate (Or) at (5,0);
		\coordinate (Ol) at (-2,0);
		\coordinate (S) at (-0.5,0);
		\coordinate (E) at (2.5,0);
		\coordinate (H) at ($({2.5+1.5*cos(60)},{1.5*sin(60)})$);
		\coordinate (P) at
		($({2.5+1.5*cos(60)+0.7*cos(60+140)},{1.5*sin(60)+0.7*sin(60+140)})$);
		\coordinate (Q) at ($({2.5+2.4*cos(60)},{2.4*sin(60)})$);
		\coordinate (R) at ($({2.5+1.5*cos(35)},{-1.5*sin(35)})$);
		\coordinate (R1) at ($({2.5+0.75*cos(35)},{-0.75*sin(35)})$);
		\coordinate (F) at
		($({2.5+1.5*cos(60)+0.9*cos(30)},{1.5*sin(60)-0.9*sin(30)})$);
		\coordinate (G) at
		($({2.5+1.5*cos(60)-0.9*cos(30)},{1.5*sin(60)+0.9*sin(30)})$);

		\draw (O) node [right=0.2cm,below] {$O$};
		\draw [-latex] (Ol)--(Or) node [below] {$x$};
		\draw [-latex] (Od)--(Ou) node [left] {$y$};
		
		\draw (S) circle (0.6mm) node [left=0.5mm,below] {$S$};
		\fill (S) circle (0.6mm);
		
		\draw (E) circle (1.5) node [left=0.5mm,below] {$E$};
		\fill (E) circle (0.6mm);
		
		\draw (H) circle (0.7) node [right=0.3mm,above=0.25cm] {$\delta$};
		\fill (H) circle (0.6mm);
		
		\draw (R1) node [below] {$d_0$};
		
		\draw (P) node [left] {$v_0$};
		
		\pic [draw, "$\,\,\beta$", angle eccentricity=1.4, angle radius =
		0.4cm] {angle = Or--E--H};
		\pic [draw, angle eccentricity=1.4, angle radius = 0.3cm] {angle
			= Q--H--P};
		
		\draw [dashed] (H)--(Q);
		
		\draw [-latex] (E)--(H);
		\draw [-latex] (H)--(P);
		\draw [dashed] (E)--(R);
		\draw [dashed] (F)--(G);
		\end{tikzpicture}
	\end{center}
	\caption{In the synodic barycentric reference frame, the
		initial conditions are defined using two angles: $\beta$
		gives the position of the small body with respect to the
		Earth, $\delta$ gives the direction of the velocity.}
	\label{fig:synInCond}
\end{figure}
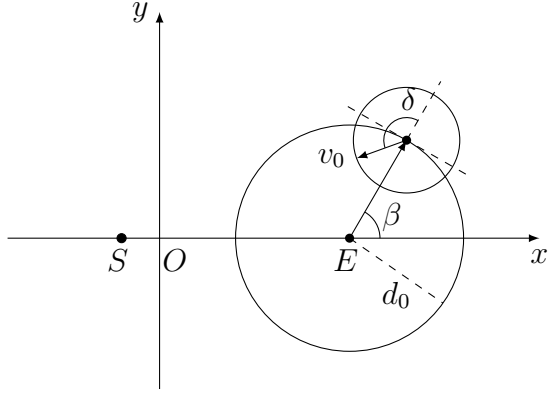

Let us fix the value $C$ of the Jacobi integral $\jacobi$ and the initial
distance $d_0$ between the particle and the Earth. We select $d_0$ such that the variation of the Tisserand parameter $\vert \frac{dT}{dt}(d_0)\vert$ is small. In this way,  at the initial time $t_0$, the motion of the particle can be approximated by a Keplerian orbit around the Sun. We choose
\begin{equation}
d_0 = (1+\eps) \cdot d_C,
\label{d0def}
\end{equation}
with $d_C$ defined in Section \ref{sec:tiss} and $0<\eps\ll 1$; $\eps$ is set in order to have $\vert
\frac{dT}{dt}(d_0)\vert<5\cdot10^{-4}$ for all $C\in[-3,3]$ (see relation \ref{dTdtMinEps}).
The initial positions of the particle are selected on the circle
\begin{equation}
\mathcal{D}_0=\{(x,y)\in\mathcal{R}: (x-1+\mu)^2+y^2=d_0^2\}.
\end{equation}
Given an initial position $(x_0,y_0)=\big(x(t_0),y(t_0)\big)$, the
norm of the velocity $v_0$ is
completely determined by the value of the Jacobi integral $C$. Indeed, from
equation \eqref{Cdef}, we have
\begin{equation}
v_0^2 = \dot{x_0}^2+\dot{y}_0^2 =
2\frac{1-\mu}{r_0}+2\frac{\mu}{d_0}+x_0^2+y_0^2-C,
\label{v0def}
\end{equation}
where $r_0 = r(x_0,y_0)$, see \eqref{rdef}. We can prove that for
$C\le 3$ and for every $(x_0,y_0)\in \mathcal{D}_0$ relation \eqref{v0def} gives $v_0^2>0$. Let us set
$X=x_0-1+\mu$. We have
\begin{equation*}
v_0^2 = \tildeg(X) + 2\frac{\mu}{d_0} + d_0^2 + (1-\mu)^2 - C,
\end{equation*}
where
\begin{equation*}
\tildeg(X)=2\frac{1-\mu}{\sqrt{1+d_0^2+2X}}+2(1-\mu)X.
\label{v0max_gfun}
\end{equation*}
The function $\tildeg$
attains its minimum at $X=-d_0^2/2$. Indeed,
\begin{equation*}
\frac{\mathrm{d} \tildeg}{\mathrm{d} X }(X)=
2(1-\mu)\left(1-\left(1+d_0^2+2X\right)^{-3/2}\right)
\end{equation*}
and 
\begin{equation*}
\frac{\mathrm{d}^2 \tildeg}{\mathrm{d}X^2}
(X)=\frac{6(1-\mu)}{\left(1+d_0^2+2X\right)^{5/2}}>0 \qquad
\mbox{ for }|X|\leq d_0.
\end{equation*}
Thus, we get 
\[
v_0^2\ge 3-4\mu + \mu^2 + \frac{2\mu}{d_0} + \mu d_0^2-C.
\]
With the assumptions made in Section~\ref{sec:tiss} (see Fig.
\ref{fig:dref}) and from \eqref{d0def}, we have $d_0<20\rhill\sim 0.2$, so that
\begin{equation*}
v_0^2 > 3-C.
\end{equation*}
We can parametrize the initial conditions with two angular coordinates
$\beta$ and $\delta$. Let us consider a geocentric synodic reference
frame $Ex_Py_Pz_P$ with the same orientation of $Oxyz$: $\beta$ is the
angle between the $x_P$-axis and the initial position of the particle,
$\delta$ is the angle between the initial position of the particle and
its velocity, see Fig. \ref{fig:synInCond}. The initial conditions
in barycentric synodic coordinates are given by
\begin{equation*}
\begin{aligned}
x_0&=1-\mu+d_0\cos\beta,&\qquad \dot x_0&=v_0\cos(\beta+\delta),\\
y_0&=d_0\sin\beta,&\qquad \dot y_0&=v_0\sin(\beta+\delta).
\end{aligned}
\end{equation*}
Since we are interested in close encounters, the initial velocities
are selected in order to obtain trajectories entering the circle
$\mathcal{D}_0$.
This means taking $\delta\in(\frac{\pi}{2},\frac{3\pi}{2})$.
The set of initial conditions is determined by
considering a regularly-spaced grid on the $(\beta,\delta)$-plane.

\subsection{Computation of the 3-body orbit}

For each initial condition, we propagate the orbit in the
CR3BP. 
The propagation is interrupted
either when the particle reaches again the initial distance $d_0$ from
the planet
or when a maximum propagation time is reached. In the following, we denote by
$\bx^{\rm 3bp}(t)$ the orbit computed for each initial condition,
where $t\in[t_0,t_1]$ and $t_1$ is the time at which
the propagation is stopped. 

Since the particle can get very close to the Earth, the
propagation is performed using the Levi-Civita regularization
\citep{Stiefel_1971}. We introduce the variables $(p_u,p_v,u,v)$
through the canonical transformation
\begin{align}
& x = u^2-v^2+1-\mu, \\ & y = 2uv, \\ & p_x =
\frac{1}{2}\frac{p_uu-p_vv}{u^2+v^2}, \\ & p_y =
\frac{1}{2}\frac{p_uv+p_uv}{u^2+v^2}+1-\mu
\end{align}
and a fictitious time $\tau$ through relation
\begin{equation*}
\frac{\mathrm{d}\tau}{\mathrm{d}t} = \frac{1}{u^2+v^2}.
\end{equation*}
The regularized Hamiltonian is
\begin{equation}
\begin{split}
K & =  \frac{1}{8}\,\Big(p_u+2v(u^2+v^2)\Big)^2
+\frac{1}{8}\,\Big(p_v-2u(u^2+v^2)\Big)^2 \\ & \quad
-(1-\mu)(u^2+v^2)\left(\frac{1}{\sqrt{1+2(u^2-v^2)+\vert u\vert
		^4}}-1+u^2-v^2\right) \\ & \quad -\frac{1}{2}\,(u^2+v^2)^3
-(u^2+v^2)\left(E+1-\mu+\frac{(1-\mu)^2}{2}\right)-\mu,
\end{split}
\label{eqn:LCham}
\end{equation}
where $E$ is the constant value of the Hamiltonian function $H_R$ in
\eqref{3bpHam} evaluated at the initial conditions
$({x}_0,{y}_0,\dot{x}_0,\dot{y}_0)$.

The equations of motion are then
\begin{equation*}
\qquad p_u' = -\frac{\partial K}{\partial u},\quad p_v' =
-\frac{\partial K}{\partial v}, \quad u' = \frac{\partial K}{\partial
	p_u},\quad v' = \frac{\partial K}{\partial p_v},
\end{equation*}
with the primed quantities corresponding to the derivatives with
respect to the fictitious time $\tau$.

\subsection{Optimisation process}

For each initial condition and given the associated 3-body orbit
$\bx^{\rm 3bp}(t)$, we search for $\Rsel\in \mathcal{D}_{\rsoivar}$ minimizing the
target function $f:\mathcal{D}_{\rsoivar}\mapsto \mathbb{R}$ defined as
\begin{equation}
f(\rsoivar) =  \sup_{t\in[t_0,t_1]}\vert \bx^{\rm 3bp}(t)-\bx^{\rm
	pc}(t;\rsoivar) \vert + \vert \bx^{\rm 3bp}(t_1)-\bx^{\rm
	pc}(t_1;\rsoivar) \vert + \vert \bx^{\rm 3bp}(t_{\q3bp})-\bx^{\rm
	pc}(t_{\qpcm};\rsoivar) \vert
\label{optnorm}
\end{equation}
where $\bx^{\rm pc}(t;\rsoivar)$ is the orbit obtained by
applying the patched-conic method with a sphere of influence of radius
$\rsoivar$ and $t_{\q3bp}$ and $t_{\qpcm}$ are respectively the times
of minimum geocentric distance of the 3-body orbit and of the
patched-conic orbit. Let us remark that $t_{\q3bp}$ and $t_{\qpcm}$
are generally different.
Because of the three different terms of $f(\rsoivar)$, the optimisation procedure allows us to determine a patched-conic orbit with the following features: it is close to the 3-body orbit in the phase space over the whole time interval $[t_0,t_1]$; its final state is similar to the final state of the 3-body orbit; its state at the perigee is similar to the state of the 3-body orbit at $t=t_{\q3bp}$. 
The domain of $f(\rsoivar)$ is defined as
\begin{equation*}
\mathcal{D}_{\rsoivar}=\{\rsoivar\in\mathbb{R}:\max(\radEarth,\q3bp)\leq
\rsoivar\leq \Rmax \}.
\end{equation*}
We remind that $\q3bp$ is the minimum geocentric distance reached
by the particle along the 3-body trajectory. The upper bound $\Rmax$ is chosen as a distance from the Earth where the gravitational influence of the
planet is negligible with respect to that of the Sun. In our
computation, we set $ \Rmax=5.5\rhill$ so that the ratio between the magnitudes of the gravitational attraction of the Earth and the Sun is
\begin{equation*}
\frac{\frac{\mu}{\Rmax^2}}{\frac{1-\mu}{r^2}}\leq
\frac{\mu}{1-\mu} \frac{(1+ \Rmax)^2}{ \Rmax^2} \simeq
1.1\cdot 10^{-3}.
\end{equation*}
If $\q3bp\ge \Rmax$, we assume that there is no close
encounter, thus $\bx^{\rm 3bp}(t)$ can be
entirely approximated by a Keplerian heliocentric orbit. We
summarise the steps performed to determine $\Rsel$:
\begin{enumerate}
	\item sample the domain $\mathcal{D}_{\rsoivar}$ with $m$ equi-spaced
	points ${\rsoivar}_{_i}$,  for $i= 1,\ldots,m$;
	\item compute $\bx^{\rm pc}(t;{\rsoivar}_{_i})$ for each $i$. 
	\item evaluate $f({\rsoivar}_{_i})$ and identify the subset
	$I$ of $\mathcal{D}_{\rsoivar}$ containing the minimum point of
	$f(d)$;
	\item re-sample the interval $I$ with $n$ points ${\rsoivar}_{_k}$ for
	$k= 1,\dots,n$;
	\item compute $\bx^{\rm pc}(t;{\rsoivar}_{_k})$, and
	evaluate $f({\rsoivar}_{_k})$ for all $k$. Comparing the results,
	choose $\Rsel$ such that
	\begin{equation}
	f(\Rsel) = \min_{k\in\{1,\ldots,n\}}f({\rsoivar}_{_k}).
	\end{equation}
\end{enumerate}
It is important to remark that, for some $d_i$ in step (i) above, we could
obtain a trajectory which does not enter the sphere of influence:
in that case, $\bx^{\rm pc}(t;{\rsoivar}_{_i})$ corresponds to a
Keplerian heliocentric orbit. The same can occur for some $d_k$ in step (v). 
Moreover, we are not interested
in encounters that last a very short time: {in these
	cases, the patched-conic method would give small corrections
	which can be neglected}. Thus, we avoid the orbit patching and
use a purely heliocentric orbit if
\begin{equation}
|\fPpcm|< \fPpcmmin.  
\label{trueanomthres}
\end{equation}
at the entrance of the sphere of
influence. We use $\fPpcmmin=5\, \mbox{degrees}$: this choice will be discussed
in Section~\ref{sec:results}.

During the procedure, if we identify different local minimum points of
$f(\rsoivar)$ with the same minimum value, we select the smallest
one. Furthermore, to reduce the computational cost, if for some $k$
the value of the target function becomes larger than a certain
threshold, we avoid the computation of the values of
$f({\rsoivar}_{_j})$, for $j>k$.  In our computation we use the
threshold
\begin{equation*}
\min_{ 1\le i\le k-1}5 f({\rsoivar}_{_i}).
\end{equation*}

Finally, we check whether the selected $\Rsel$ is such that $\bx^{\rm
	pc}(t;\Rsel)$ is a purely Keplerian heliocentric orbit. In that
case, we set $\Rsel=0$.  Let us remark that the method can not be
applied if the trajectory of the particle is such that the geocentric
distance has multiple local minima lower than $\Rmax$. This phenomenon
typically occurs for high values of the Jacobi constant when the
initial osculating semi-major axis and eccentricity are approximately
close to 1 and 0, respectively, i.e. when the 3-body orbit is
close to the orbit of the Earth. Sometimes, the same phenomenon is instead
related to a gravitational capture by the Earth. In all these cases,
we discard the initial conditions, as it is not possible to apply our
procedure to compute a sphere of influence.

\subsection{Selection of the radius}
For all initial conditions such that $\Rsel=0$, the
radius of the sphere of influence $\rsoi$ is set equal to zero. Otherwise, we compute the
Keplerian heliocentric orbit $\bx^{\odot}(t)$ with
$t\in[t_0,t_1]$ and evaluate the quantity
\begin{equation}
f_{\rm KH} = \sup_{t\in[t_0,t_1]} \vert \bx^{\rm
	3bp}(t)-\bx^{\odot}(t)\vert + \vert \bx^{\rm 3bp}(t_1)-\bx^{\odot}(t_1)\vert  + \vert \bx^{\rm
	3bp}(t_{\qhelio})-\bx^{\odot}(t_{\qhelio};\rsoivar) \vert,
\label{norm_helio}
\end{equation}
where $t_{\qhelio}$ is the time of minimum distance of the small body
from the planet along the heliocentric orbit.  Then we compare the
value of $f_{\rm KH}$ with $f(\Rsel)$.  If $f_{\rm KH}\leq f(\Rsel)$
we set $\rsoi=0$, meaning that the
best approximation is given by the heliocentric orbit. Otherwise we
select $\rsoi=\Rsel$.

\subsection{Interpolation}
\label{sec:interp}
With the method described in Section~\ref{sec:newdef} we computed a
database of values of $\rsoi$ for each point of a grid in the
$(\beta,\delta)$ plane and for different values of the Jacobi constant
$C\in [C_{\rm min}, C_{\rm max}]$, with $C_{\rm min}<C_{\rm max}$. To
obtain a value of $\rsoi$ for each possible data in the domain
\begin{equation*}
{\cal D} = \Bigl\{(\beta,\delta,C): 0\leq \beta <2\pi, \frac{\pi}{2} <
\delta < \frac{3}{2}\pi, C\in [C_{\rm min}, C_{\rm max}]\Bigr\},
\end{equation*}
we apply the trilinear interpolation method described below.

Let $C_i$ be the values of $C$ for which we apply the optimisation
method described above and let $(\beta_j,\delta_k)$ be the initial
conditions we have considered.  Any point $(C,\beta,\delta)$ with
\begin{equation*}
C_i<C<C_{i+1},\quad \beta_j<\beta<\beta_{j+1},\quad
\delta_k<\delta<\delta_{k+1}
\end{equation*}
is inside the rectangular prism
\begin{equation*}
\mathscr{P} =
[C_i,C_{i+1}]\times[\beta_j,\beta_{j+1}]\times[\delta_k,\delta_{k+1}].
\end{equation*}
This method computes $\rsoi(C,\beta,\delta)$ by weighing the known
values of $\rsoi$ at the vertices of the prism $\mathscr{P}$. The
planes passing through $(C,\beta,\delta)$ and parallel to the faces of
$\mathscr{P}$, divide this prism into eight smaller ones. The value
$\rsoi^{(l)}$ at each of the eight nodes is weighted by the volume
$V_l$ of the smaller prism diagonally opposite to the node, that is
\begin{equation*}
\rsoi = \frac{1}{V_{\rm tot}}\sum_{l=1}^8\rsoi^{(l)}V_l,
\end{equation*}
where $V_{\rm tot}$ is the volume of $\mathscr{P}$.

The process is reduced to a simple bilinear or linear interpolation in
case one or two of the values $C,\beta,\delta$ coincide exactly with
the values used to generate our database.

\begin{figure}
	\centering
	\begin{subfigure}{.48\textwidth}
		\centering
		\includegraphics[width=\textwidth]{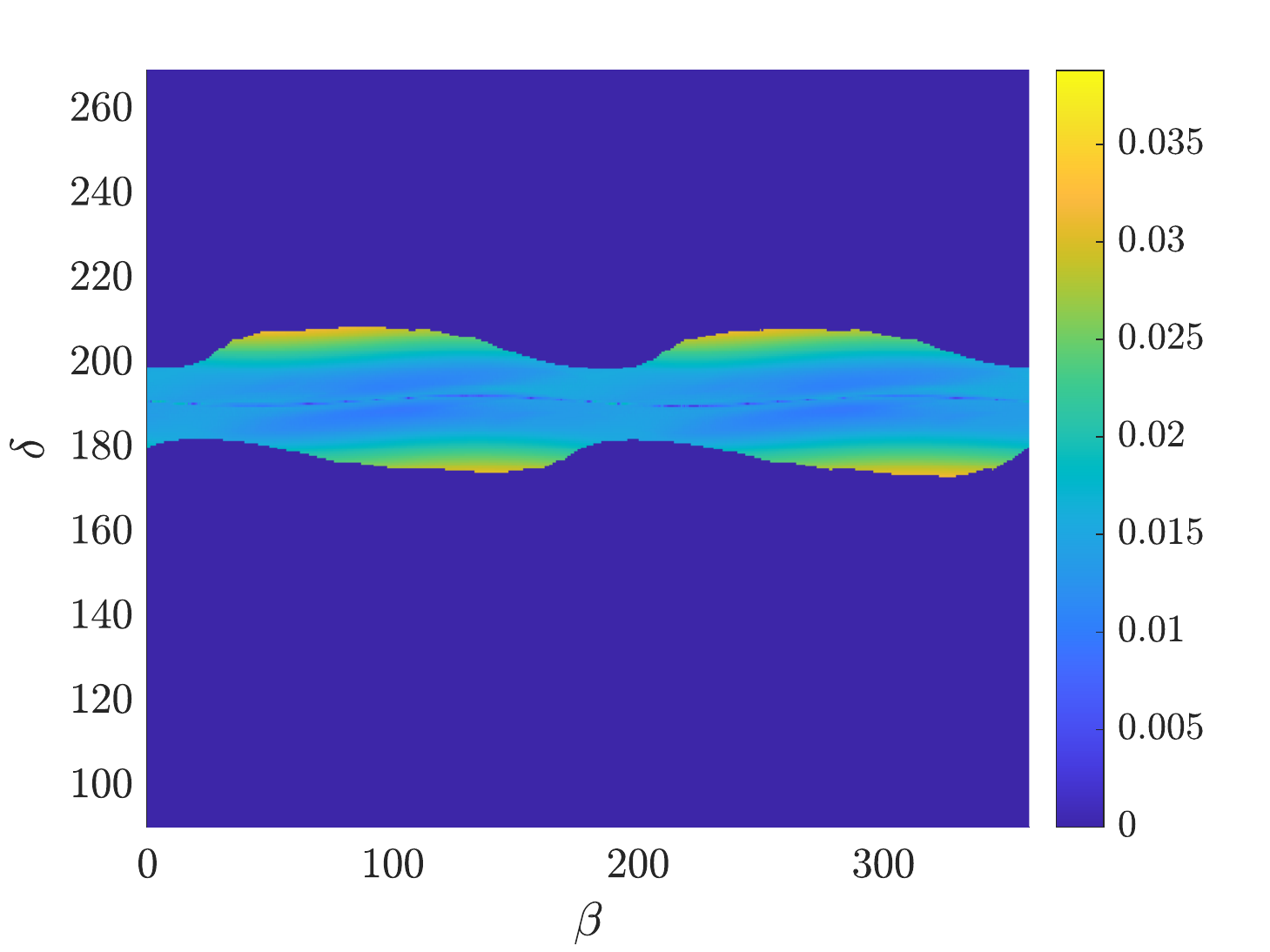}
	\end{subfigure}%
	\begin{subfigure}{.48\textwidth}
		\centering
		\includegraphics[width=\textwidth]{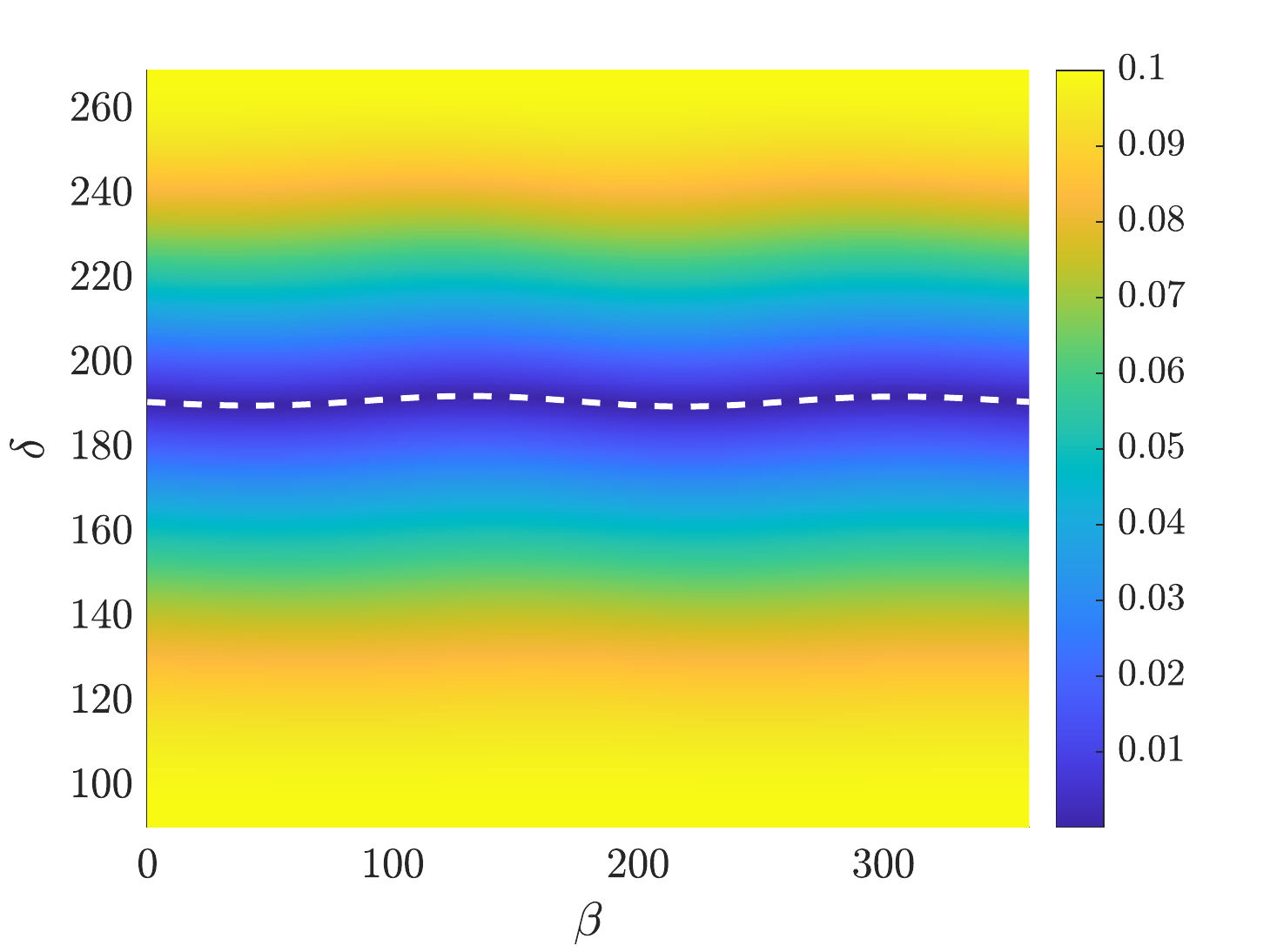}
	\end{subfigure}
	\begin{subfigure}{.48\textwidth}
		\centering
		\includegraphics[width=\textwidth]{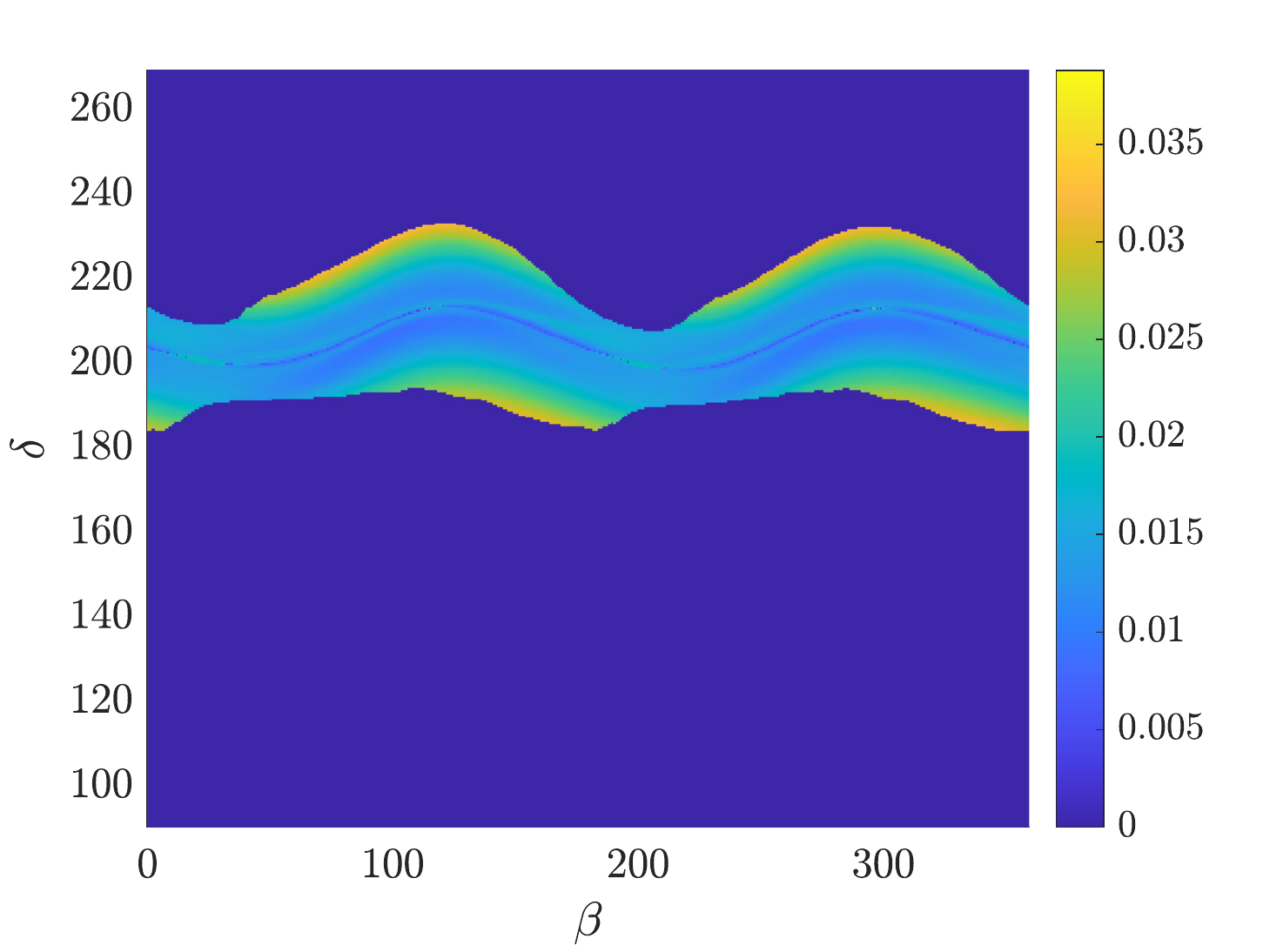}
	\end{subfigure}%
	\begin{subfigure}{.48\textwidth}
		\centering
		\includegraphics[width=\textwidth]{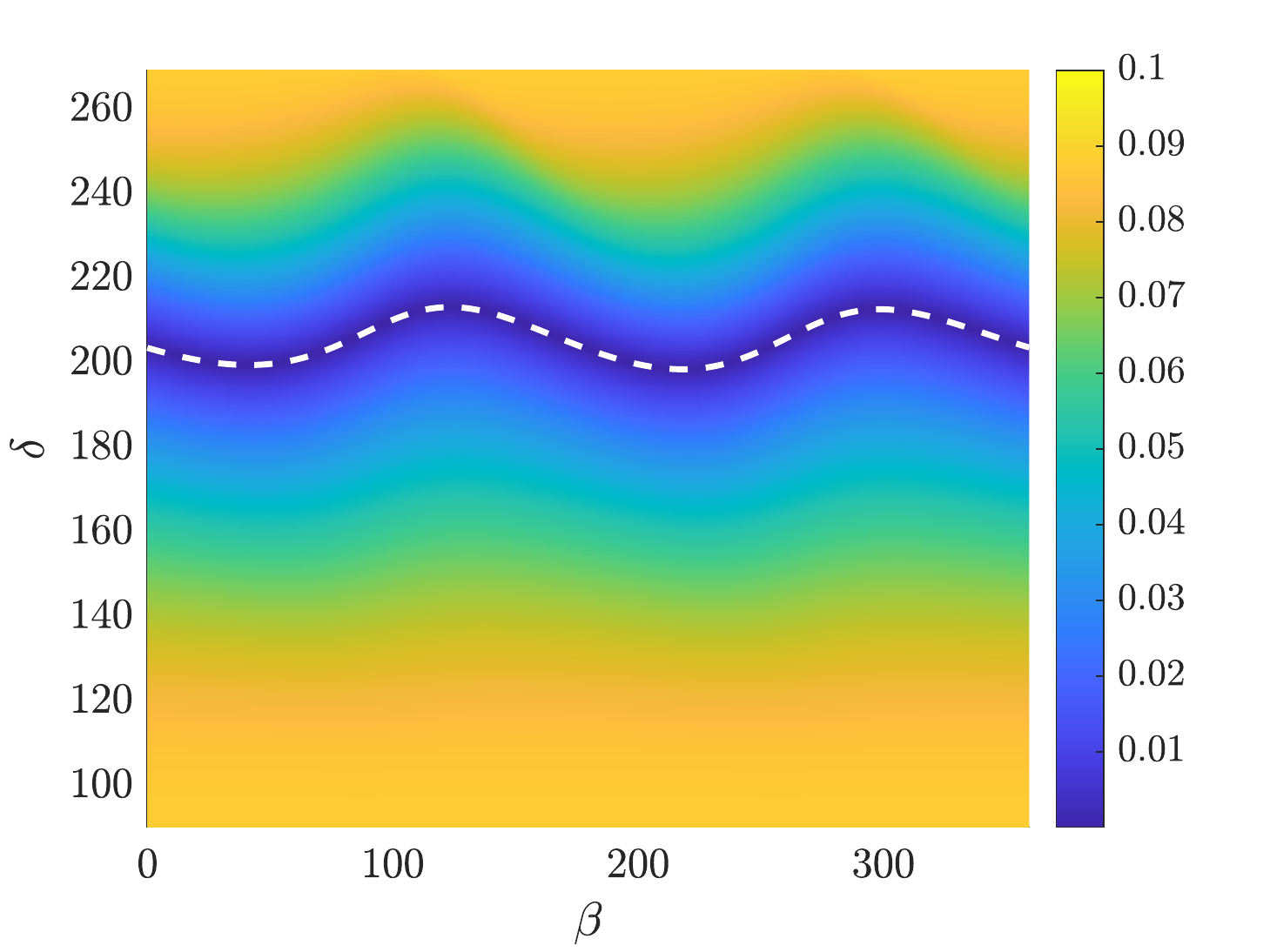}
	\end{subfigure}
	\begin{subfigure}{.48\textwidth}
		\centering
		\includegraphics[width=\textwidth]{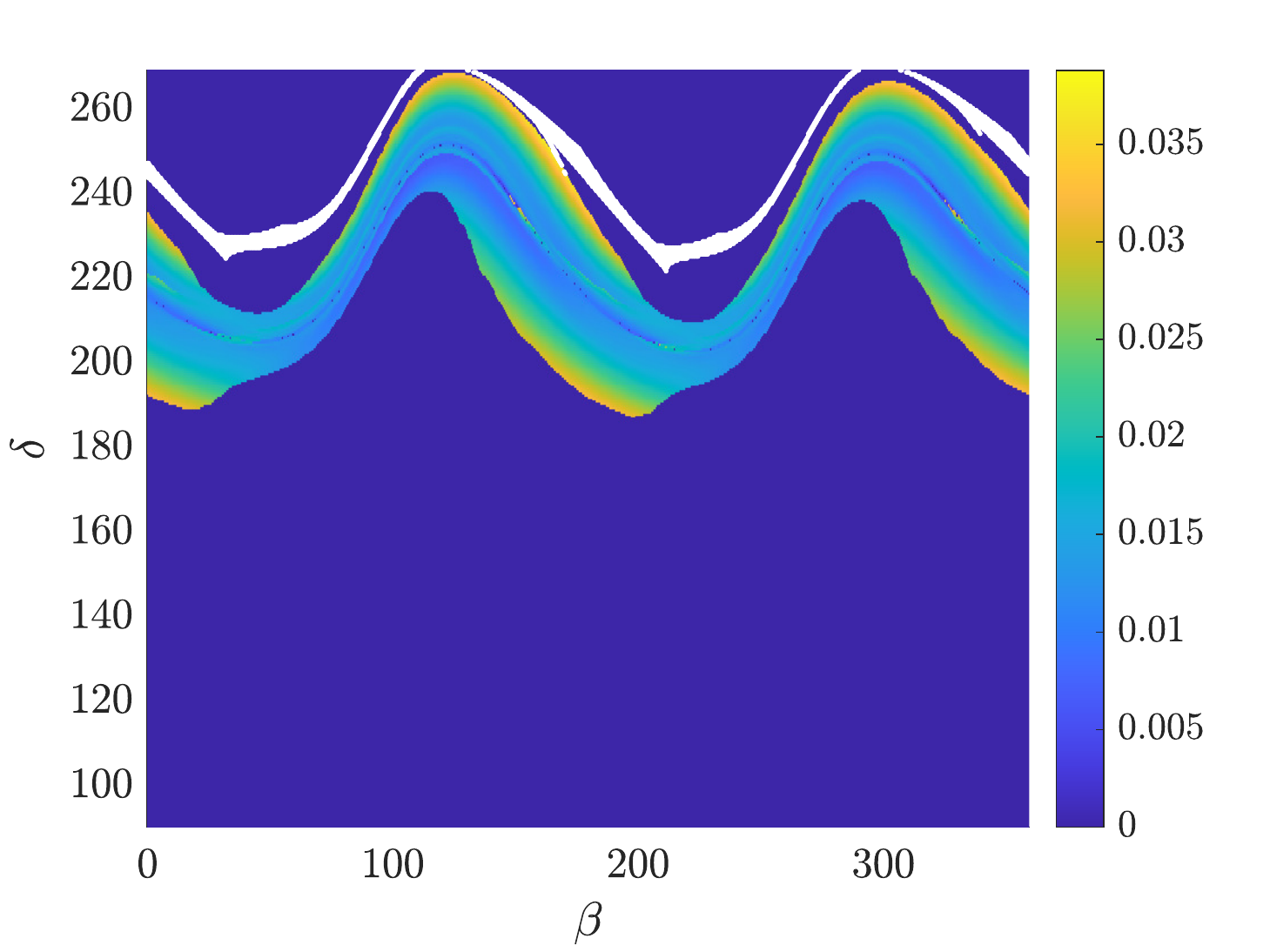}
	\end{subfigure}%
	\begin{subfigure}{.48\textwidth}
		\centering
		\includegraphics[width=\textwidth]{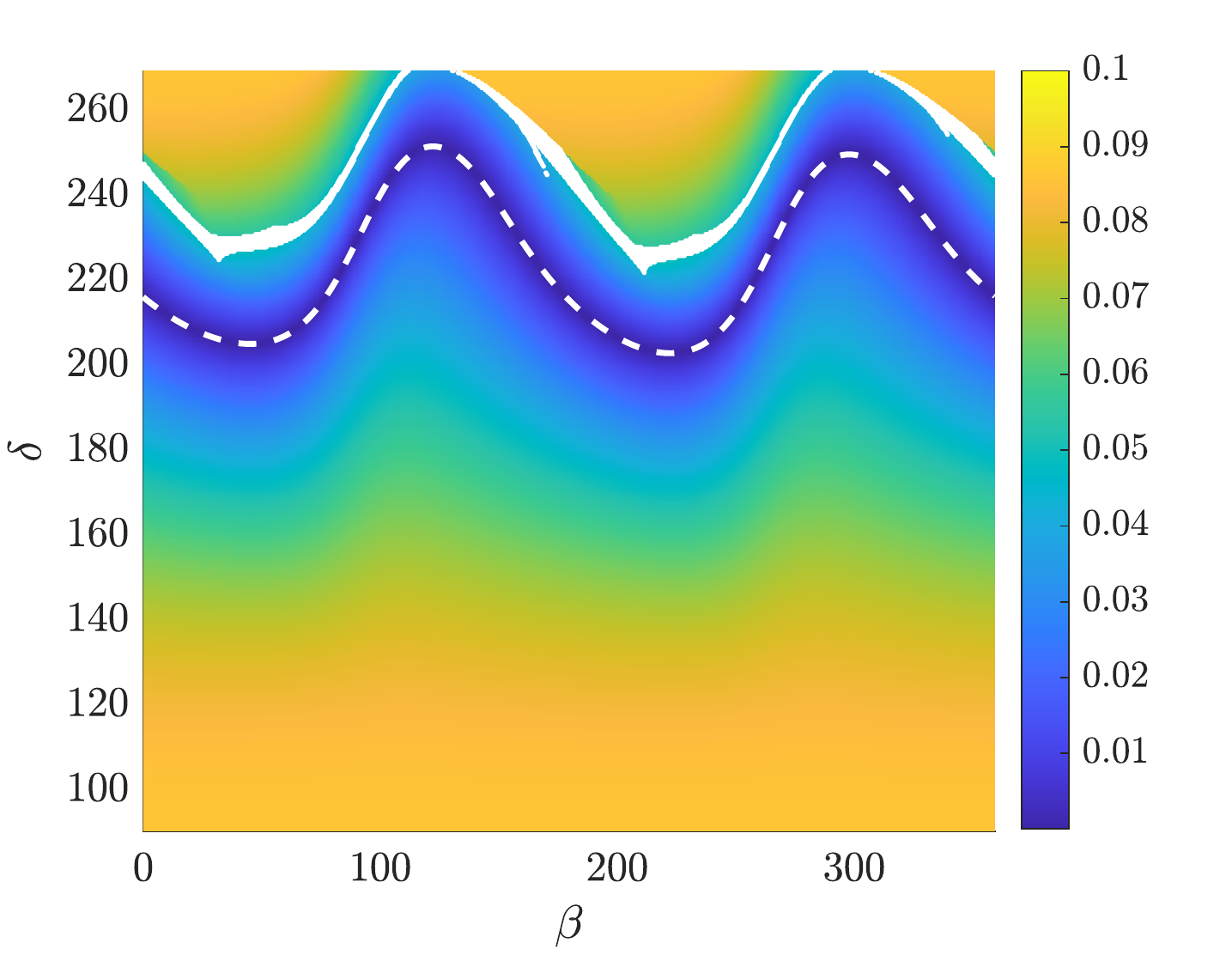}
	\end{subfigure}
	\caption{Best choice of $\rsoi$ (left) and the minimum
		distance $\q3bp$ reached during the 3-body propagation
		(right) as functions of $(\beta, \delta)$. The values of
		$\rsoi$ and $\q3bp$ are given in au, and represented with a
		colour code. Top, middle and bottom figures refer to the
		values $C=2.75, 2.97, 2.993$, respectively. The white region
		corresponds to values of $(\beta, \delta)$ for which the
		method cannot be applied.}
	\label{fig:rsoi}
\end{figure}

\begin{figure}
	\centering \centering
	\includegraphics[width=.48\textwidth]{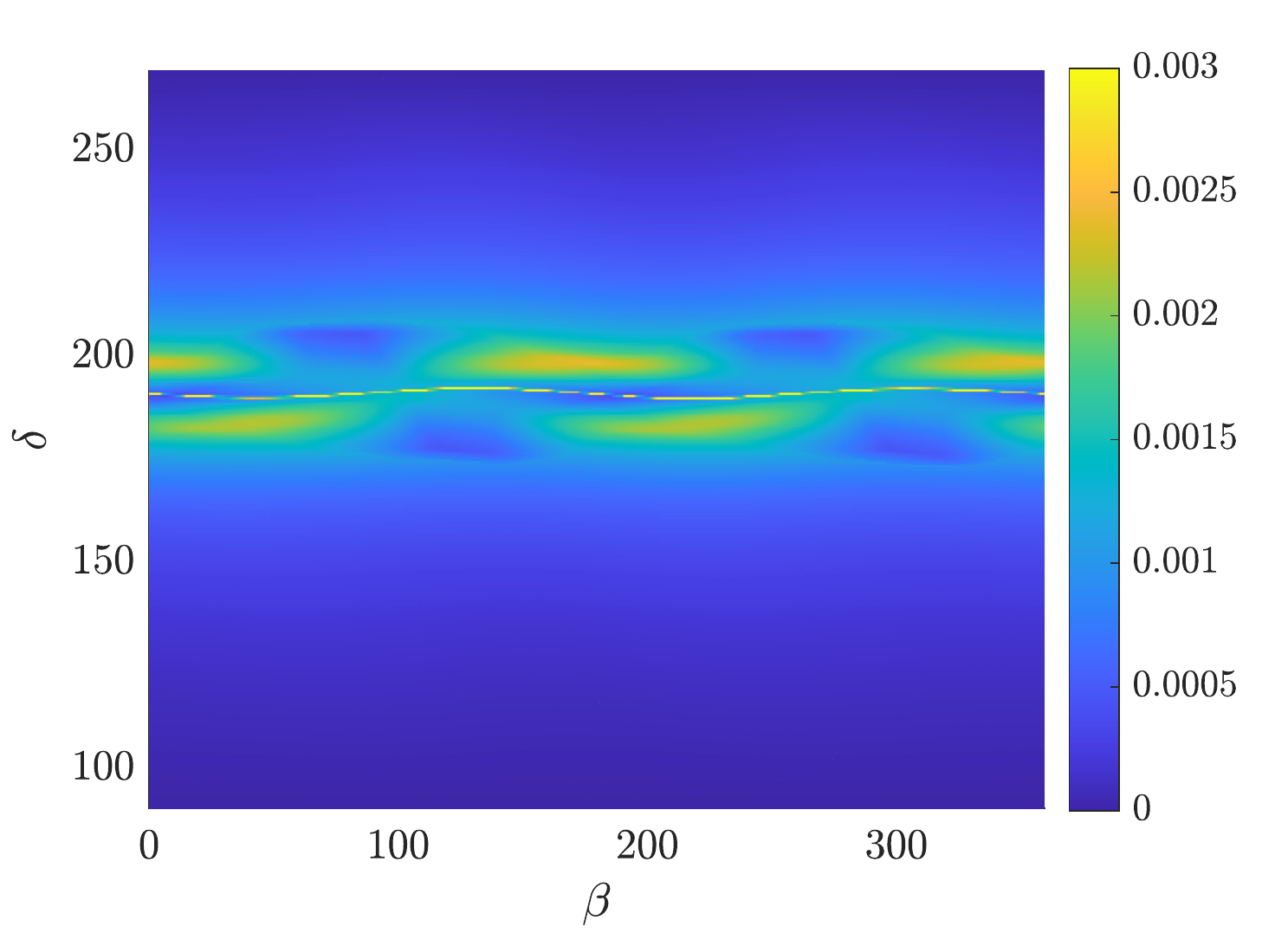}
	\\
	\includegraphics[width=.48\textwidth]{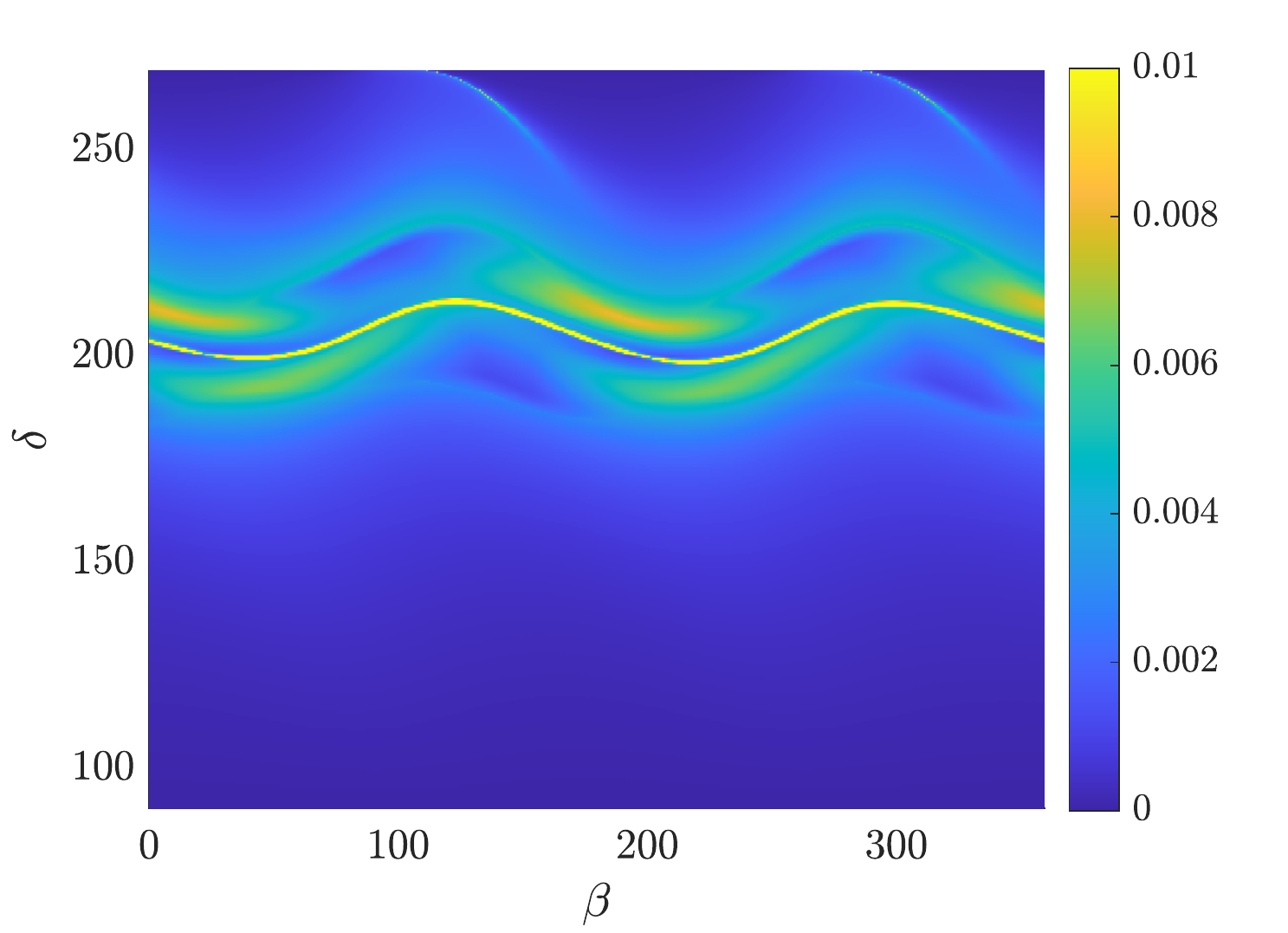}
	\\
	\includegraphics[width=.48\textwidth]{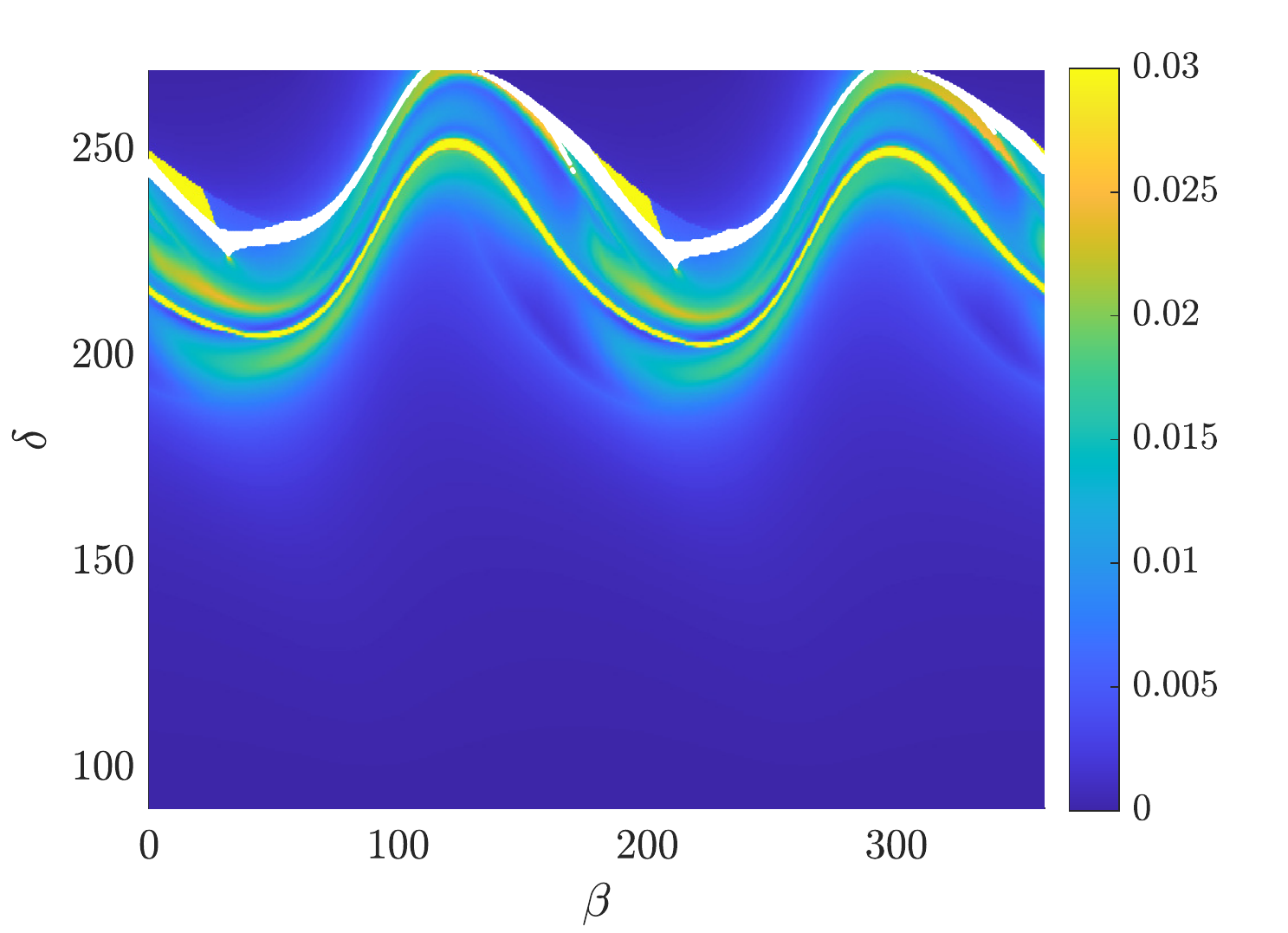}
	\caption{Values of the norm \eqref{optnorm} corresponding to
		the selected values of $\rsoi$. Top, middle and bottom figures refer to the values $C=2.75, 2.97, 2.993$,
		respectively. Note that the colour scale changes form figure to figure.}
	\label{fig:errornorm}
\end{figure}
\section{Results and discussion}
\label{sec:results}

For fixed values of the Jacobi constant, following the procedure of
the previous section, we compute $\rsoi$ for each point of a
$720\times360$ grid in the $(\beta,\delta)$ plane.  In Fig.
\ref{fig:rsoi} left, we show our results for some values of $C$,
chosen in the most significant range for near-Earth
asteroids\footnote{we used the NEODyS database {\tt https://newton.spacedys.com/neodys}}, i.e. $C\in(2,3)$. Warmer
colours correspond to larger values of $\rsoi$. The dark-blue region
corresponds to $\rsoi=0$, i.e. where a heliocentric Keplerian orbit is
chosen.
In Fig. \ref{fig:rsoi} right, we plot the values of
the minimum geocentric distance $\q3bp$ reached along the 3-body
propagation. Warmer colours correspond to larger values of
$\q3bp$. The thin, dark-blue wave represents the set of
initial conditions leading to very close encounters or
collisions. The dashed white line corresponds to
the \textit{collision} curve, i.e. the set of points $(\beta,\delta)$
for which the 3-body trajectory passes through the centre of the Earth
(in Appendix \ref{app:coll} we explain the procedure to compute it).
In the bottom images of Fig. \ref{fig:rsoi}, the white region
corresponds to points where our method cannot be applied.

Comparing the images on the right with those on the left we notice
that our method gives $\rsoi\neq0$ in the region surrounding the
collision curve. This result was expected, since this is precisely the
region where the small body gets closest to the Earth.
In Fig. \ref{fig:errornorm} we show the values of $f(\rsoi)$, defined
in \eqref{optnorm}, replaced by $f_{\rm KH}$ (see \ref{norm_helio})
if $\rsoi=0$. Warmer colours represent higher values of
$f(\rsoi)$. For the considered values of $C$, in the neighbourhood of
the collision curve, $f(\rsoi)$ achieves the highest values.
\begin{figure}[h!]
	\centering
	\includegraphics[width=.48\textwidth]{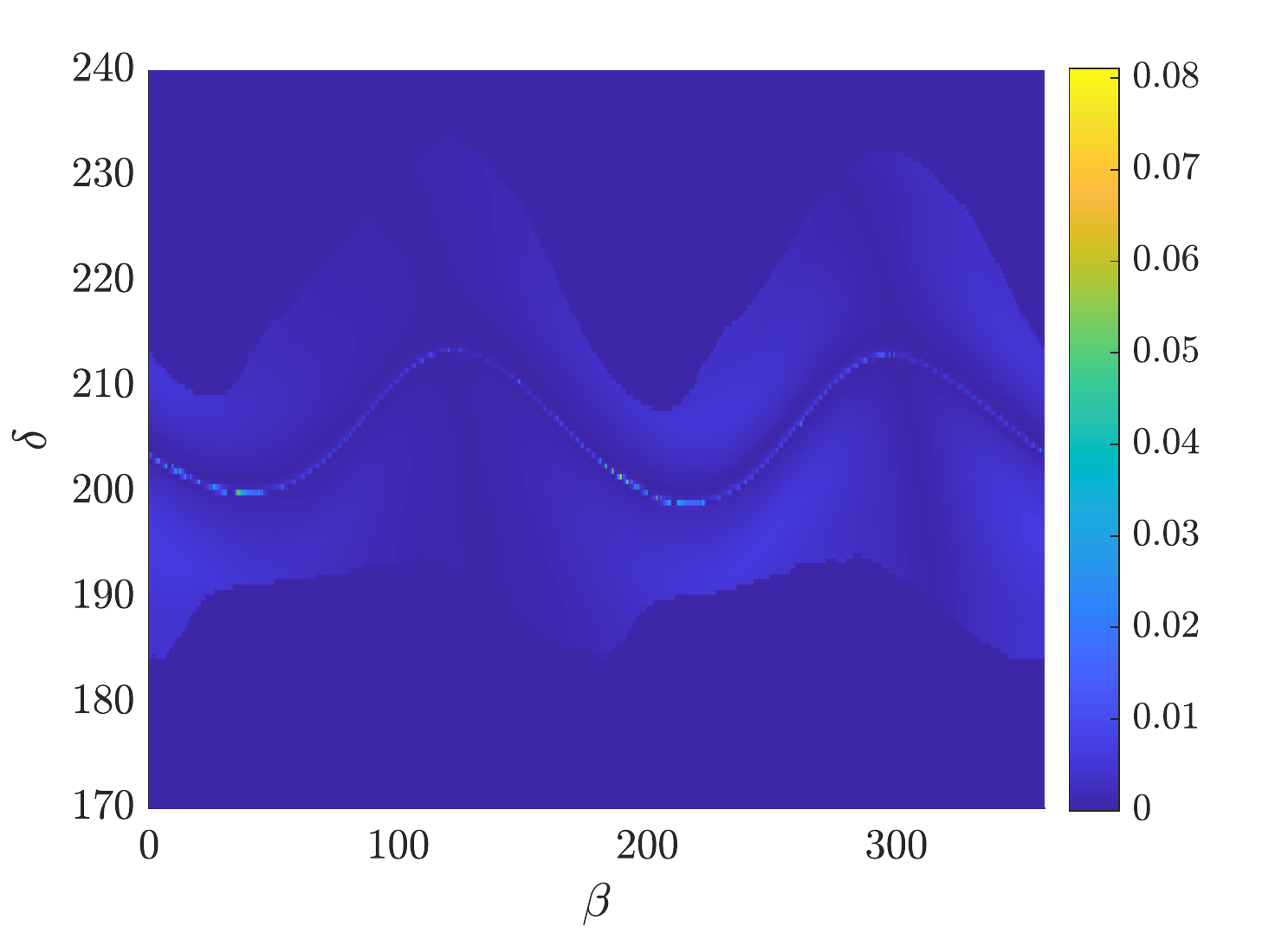}
	\includegraphics[width=.48\textwidth]{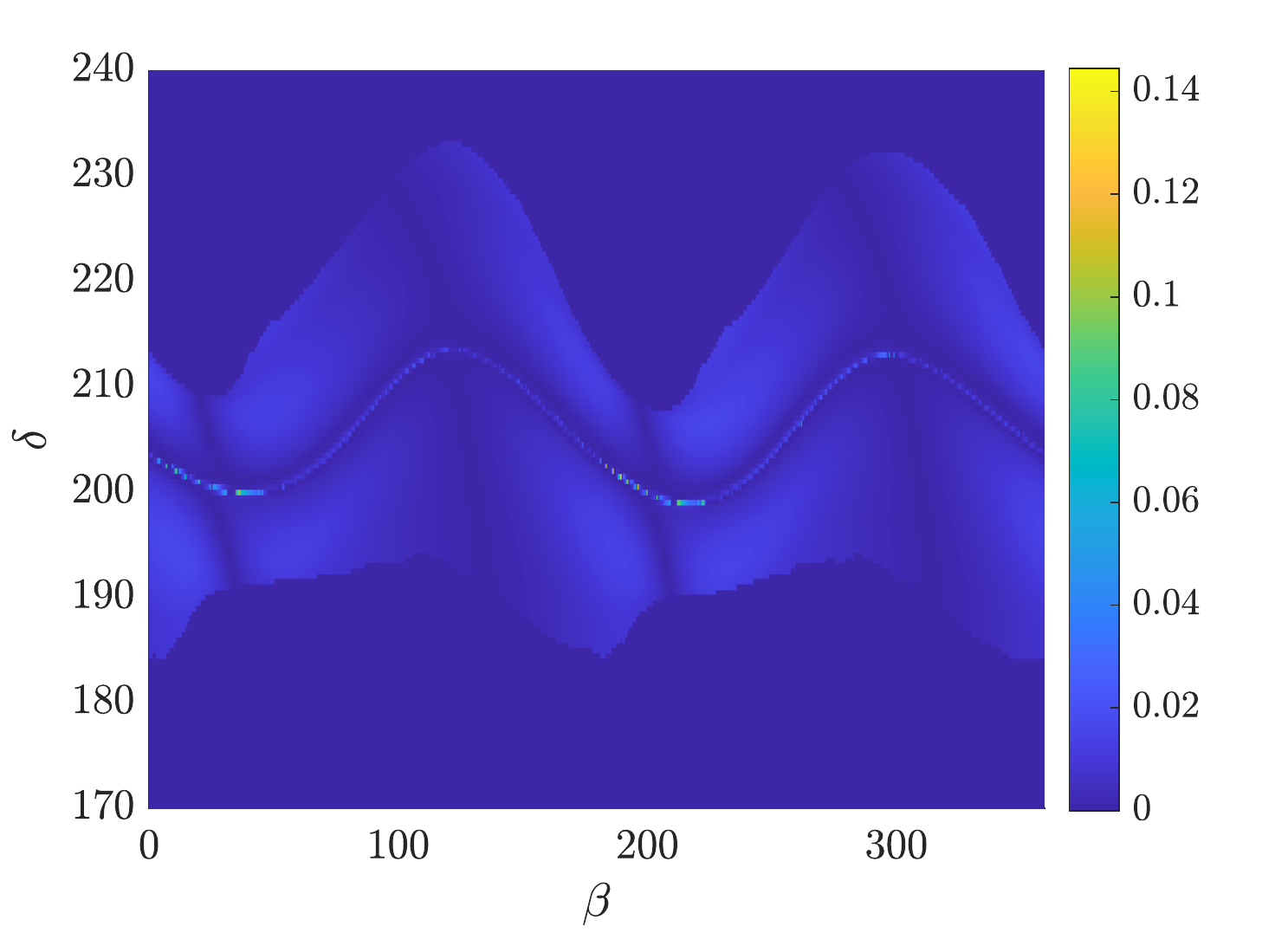}
	\caption{Relative error of the
		post-encounter semi-major axis (left) and eccentricity
		(right) for $C=2.97$.}
	\label{fig:diffSmaEccEnd}
\end{figure}

\begin{figure}[h!]
	\centering
	\includegraphics[width=.48\textwidth]{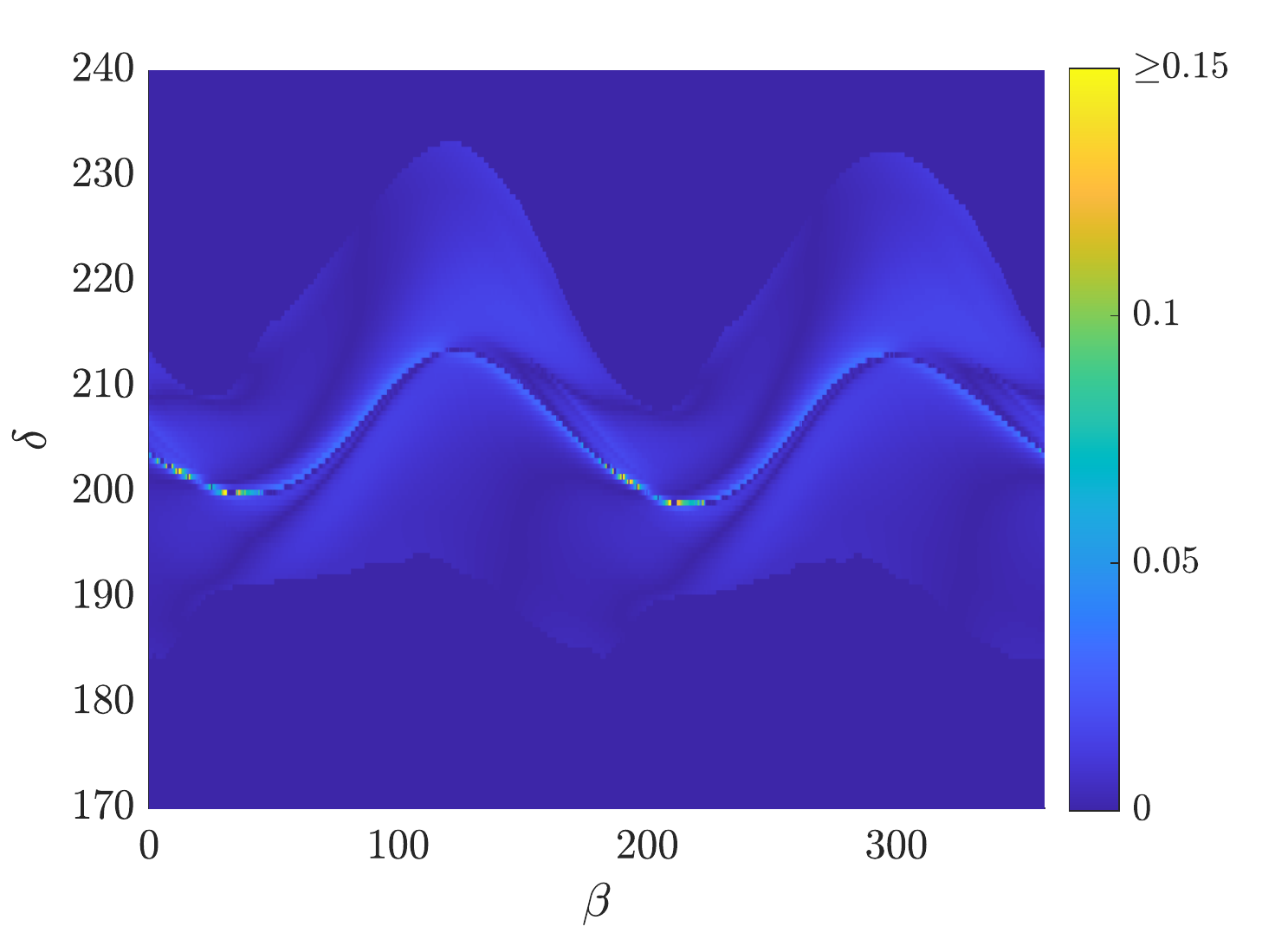}
	\includegraphics[width=.48\textwidth]{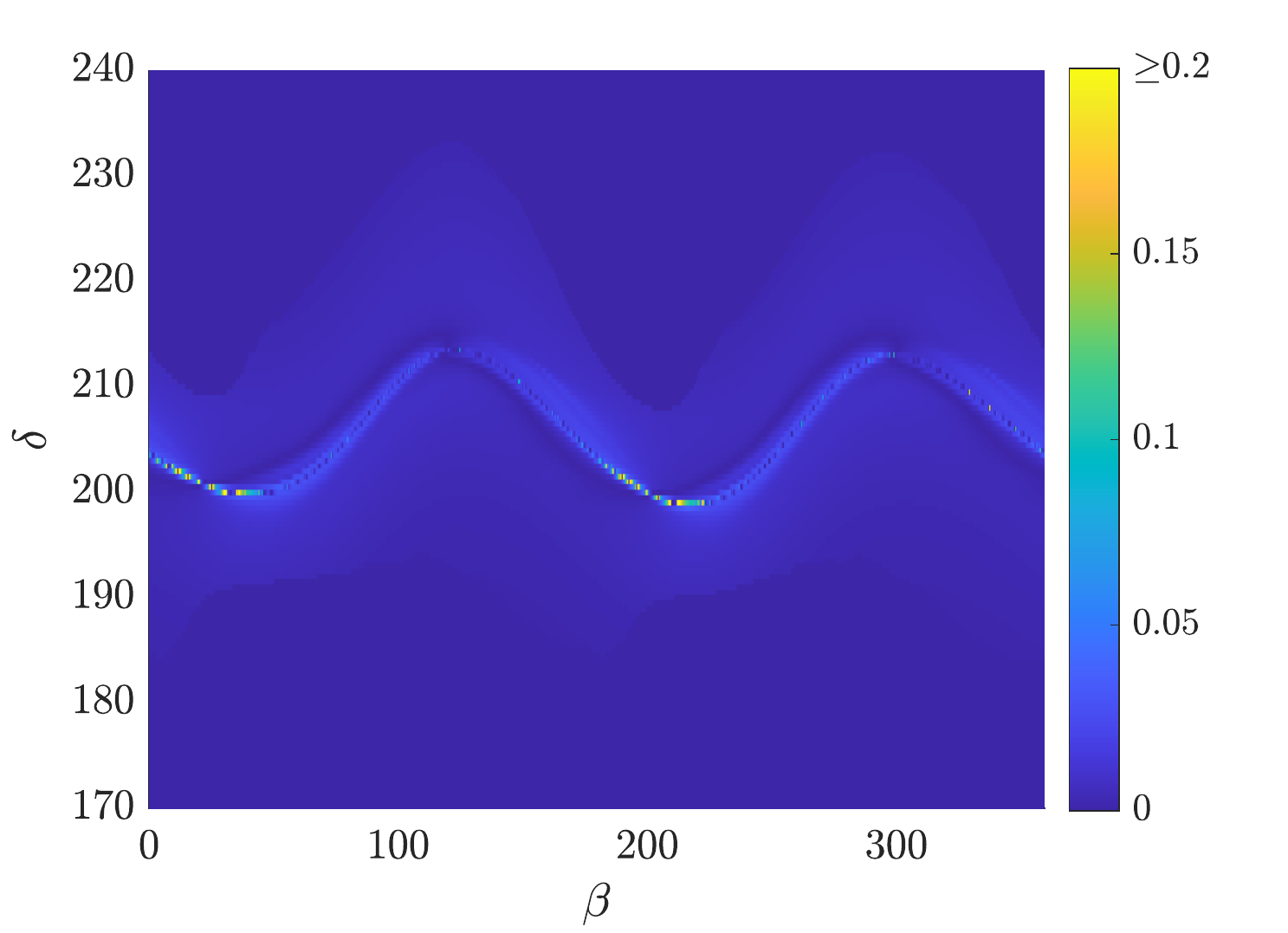}
	\caption{Relative difference between
		$\eccPpcm$ and $\eccP$ at time $t_{\q3bp}$ (left) and between the
		pericentre distances $\qpcm$ and $\q3bp$ (right) for $C=2.97$.}
	\label{fig:diffEccQ}
\end{figure}

In general, the method provides a patched-conic orbit that allows us
to reproduce some significant features of the 3-body orbit with a
reasonable error, such as the post-encounter osculating semi-major axis and
eccentricity at $t=t_1$.  To give an example, in Fig.  \ref{fig:diffSmaEccEnd} we
show the relative error of these quantities for $C=2.97$. Note that the maximum value of the relative
error is less than $9\%$ for the semi-major axis and less than $15\%$
for the eccentricity. Through our method it is usually also possible
to reproduce with sufficient accuracy the minimum geocentric distance
$\q3bp$ and the osculating geocentric eccentricity $\eccP$ at
$t_{\q3bp}$. An example can be seen in Fig.  \ref{fig:diffEccQ}
where we show the relative errors $\Delta q={\vert
	\q3bp-\qpcm\vert}/{\q3bp}$ and $\Delta e
={\vert\eccP-\eccPpcm\vert}/{\eccP}$ for $C=2.97$. For $99.97\%$ of points we have 
$\Delta e<0.15$ and for $99.94\%$ $\Delta q<0.2$. 
\begin{figure}
	\centering
	\includegraphics[width=.48\textwidth]{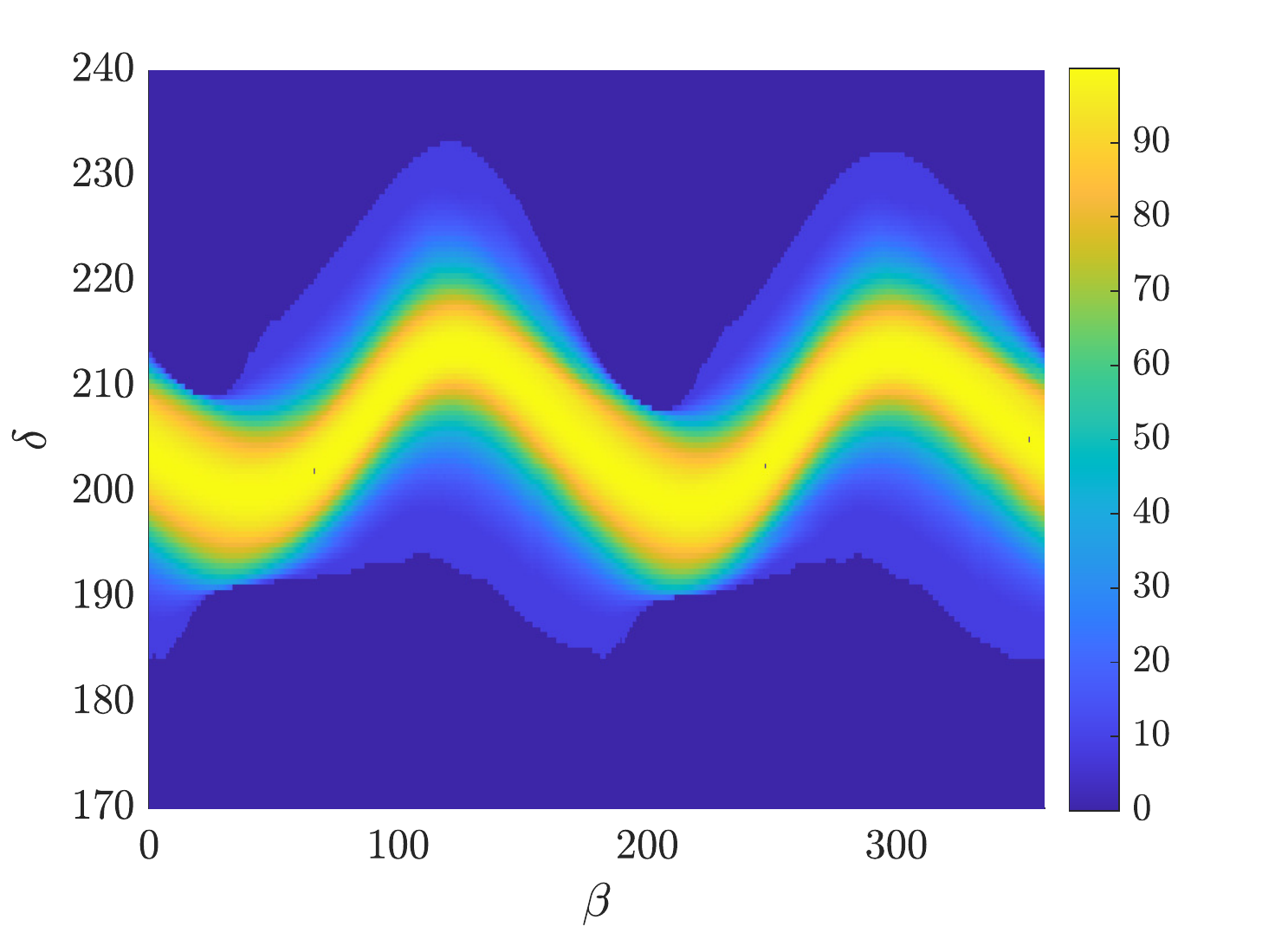}
	\caption{Percentage of deflection achieved out of the possible one. Here, $C=2.97$.}
	\label{fig:deflect}
\end{figure}
In Fig. \ref{fig:deflect} we plot, in the $(\beta,\delta)$ plane,
the percentage of achieved deflection
computed by considering $1-{\Delta\theta}/{\gamma}$, where
${\Delta\theta}/{\gamma}$ is the missed deflection, see
Section~\ref{sec:deflect}. The minimum percentage obtained in the case
considered in Fig. \ref{fig:deflect} is $8.6\%$. Although this is very
low, the percentage of achieved deflection gets higher as $\q3bp$
decreases. There is a region around the collision curve where
the achieved deflection is very close to the maximum possible one. We
argue that this is where it is important to have a high deflection
percentage. Indeed, here the planetocentric eccentricity is lower and
therefore the trajectory of the small body is significantly deflected
by the Earth. With our method we are then able to reproduce this
effect.
\begin{figure}
	\centering
	\includegraphics[width=.48\textwidth]{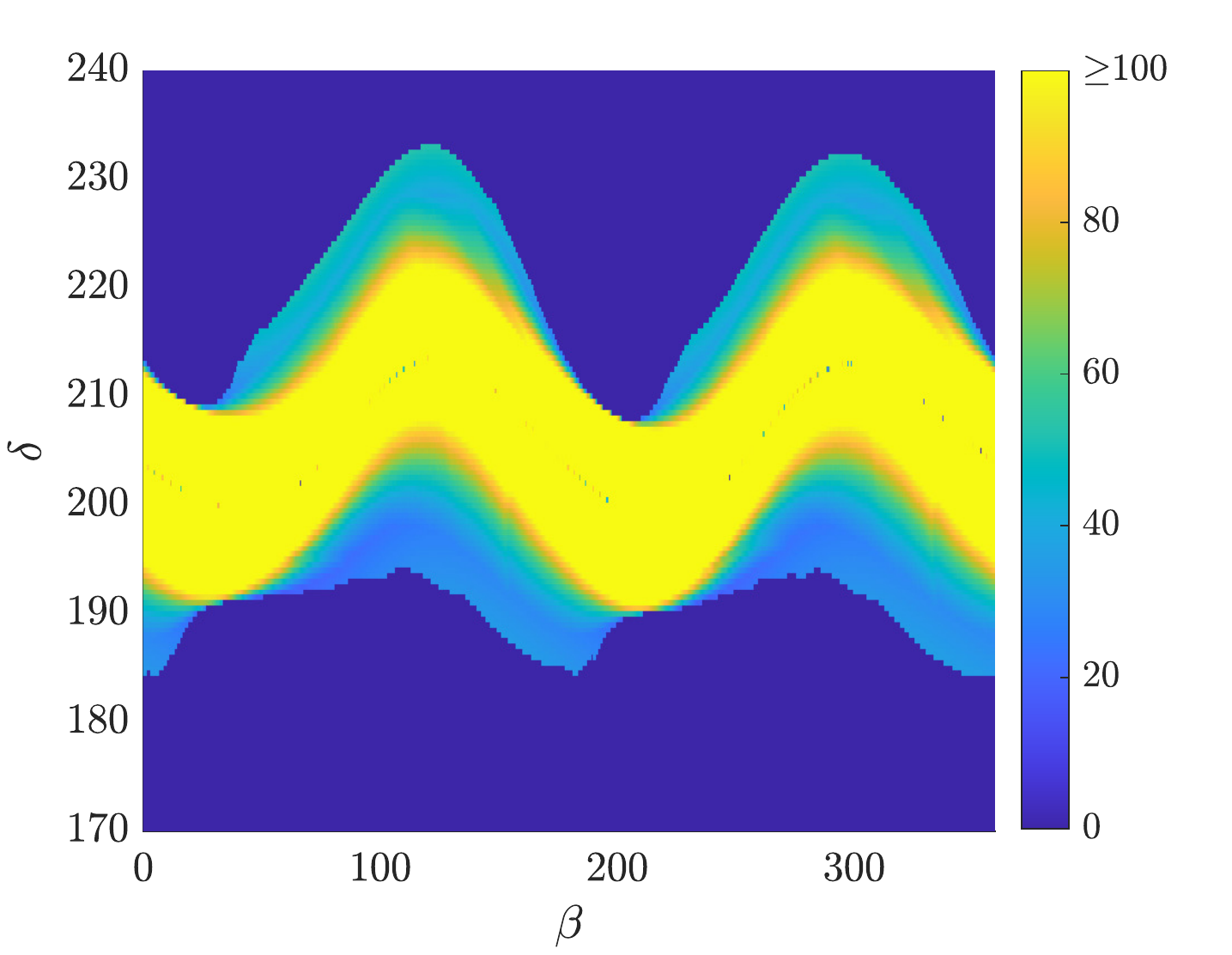}
	\includegraphics[width=.48\textwidth]{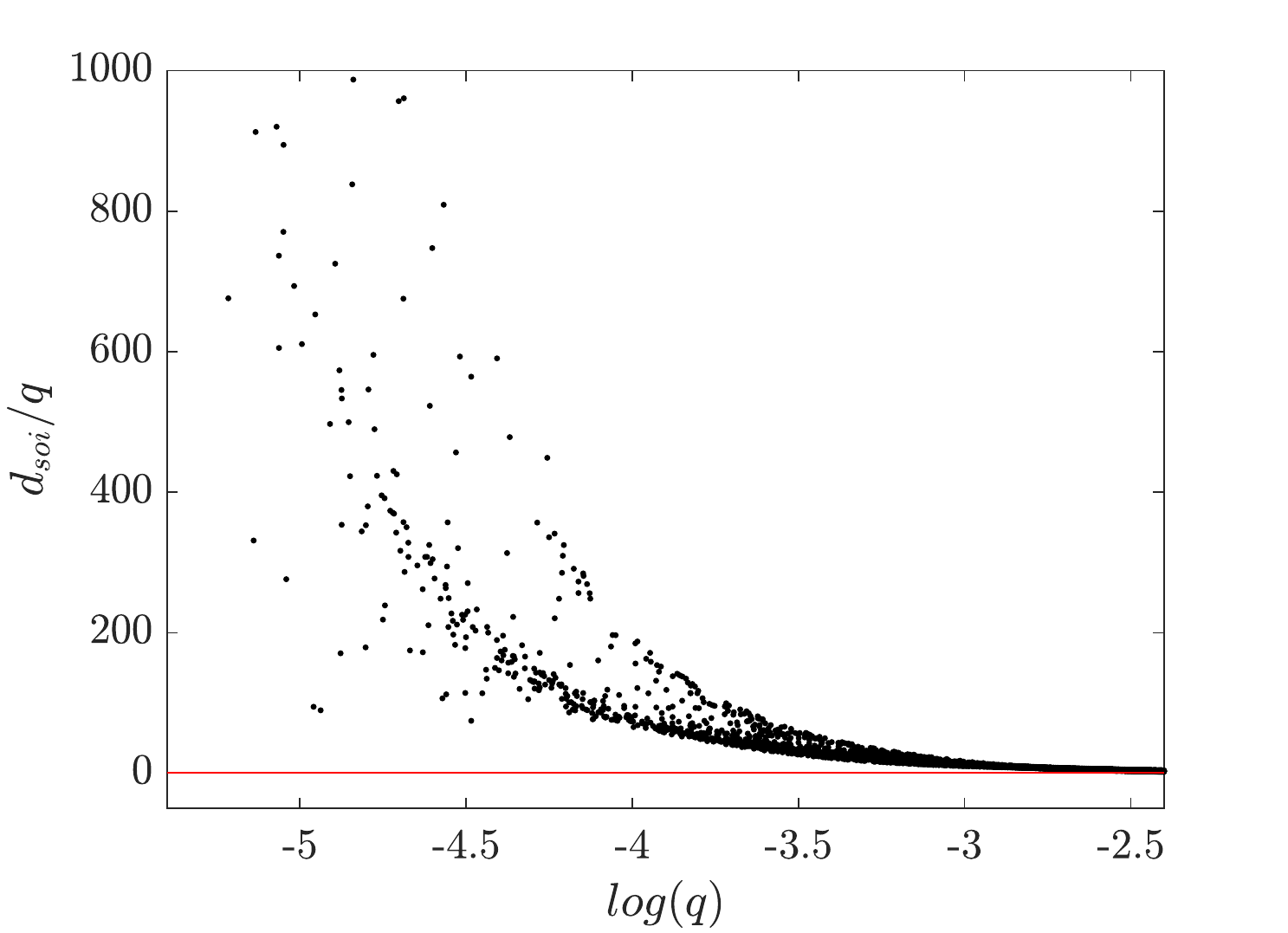}
	\caption{Left: time (in hours) spent inside the sphere of
		influence for each initial condition $(\beta,\delta)$ and
		$C=2.97$. Bright yellow points are encounters that last 100
		hours, or more. Right: ratio between $\rsoi$ and $\q3bp$. The red line corresponds to $\rsoi=\q3bp$. Note that the plot of $\rsoi/\q3bp$ presents a sort of bifurcation. This is caused by the asymmetry of $\rsoi$ with respect to the collision curve in the $(\beta,\delta)$ plane for each $C$. }
	\label{timeInRsoi}
\end{figure}

\begin{figure}
	\centering
	\includegraphics[width=.48\textwidth]{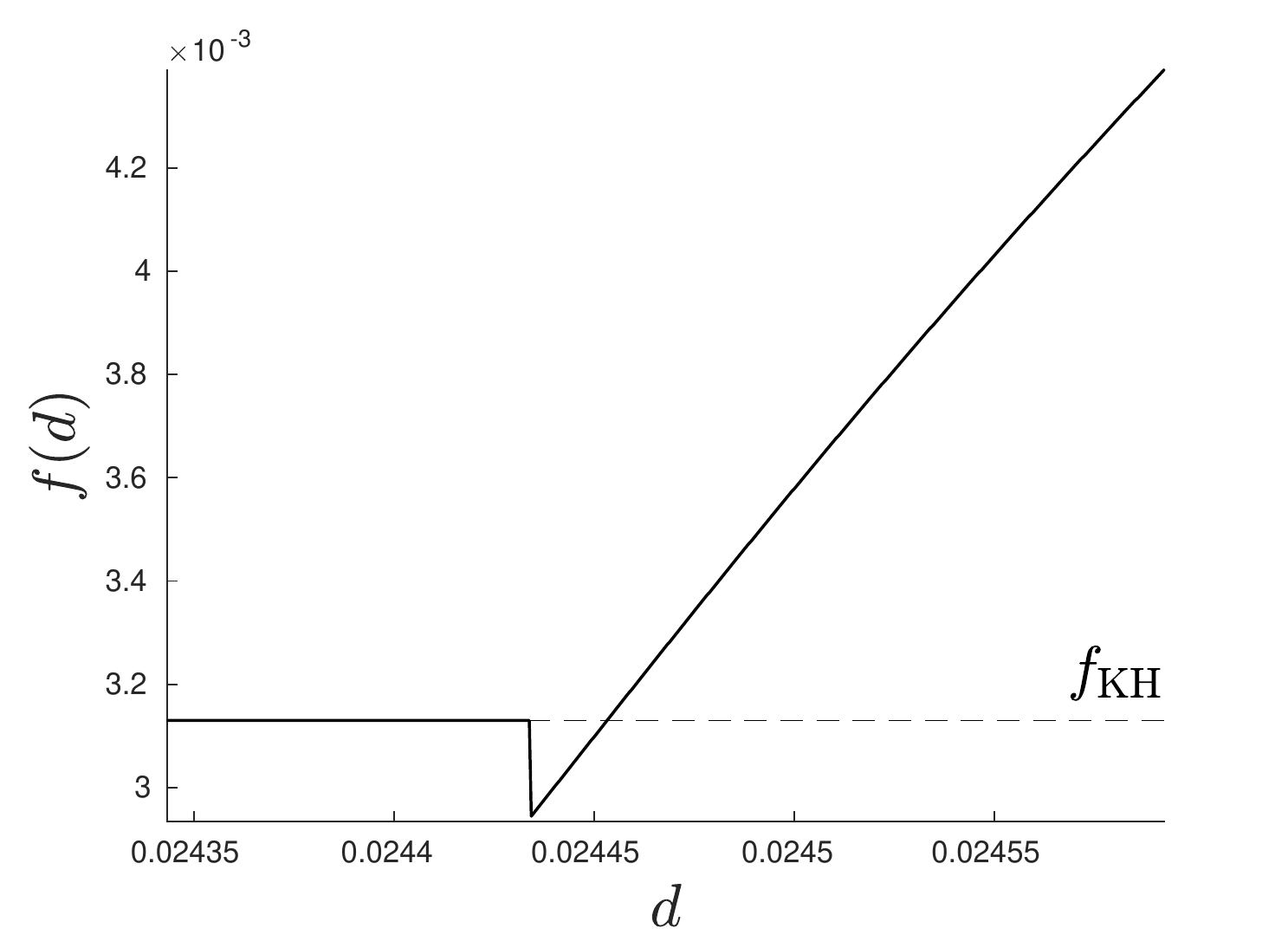}
	\includegraphics[width=.48\textwidth]{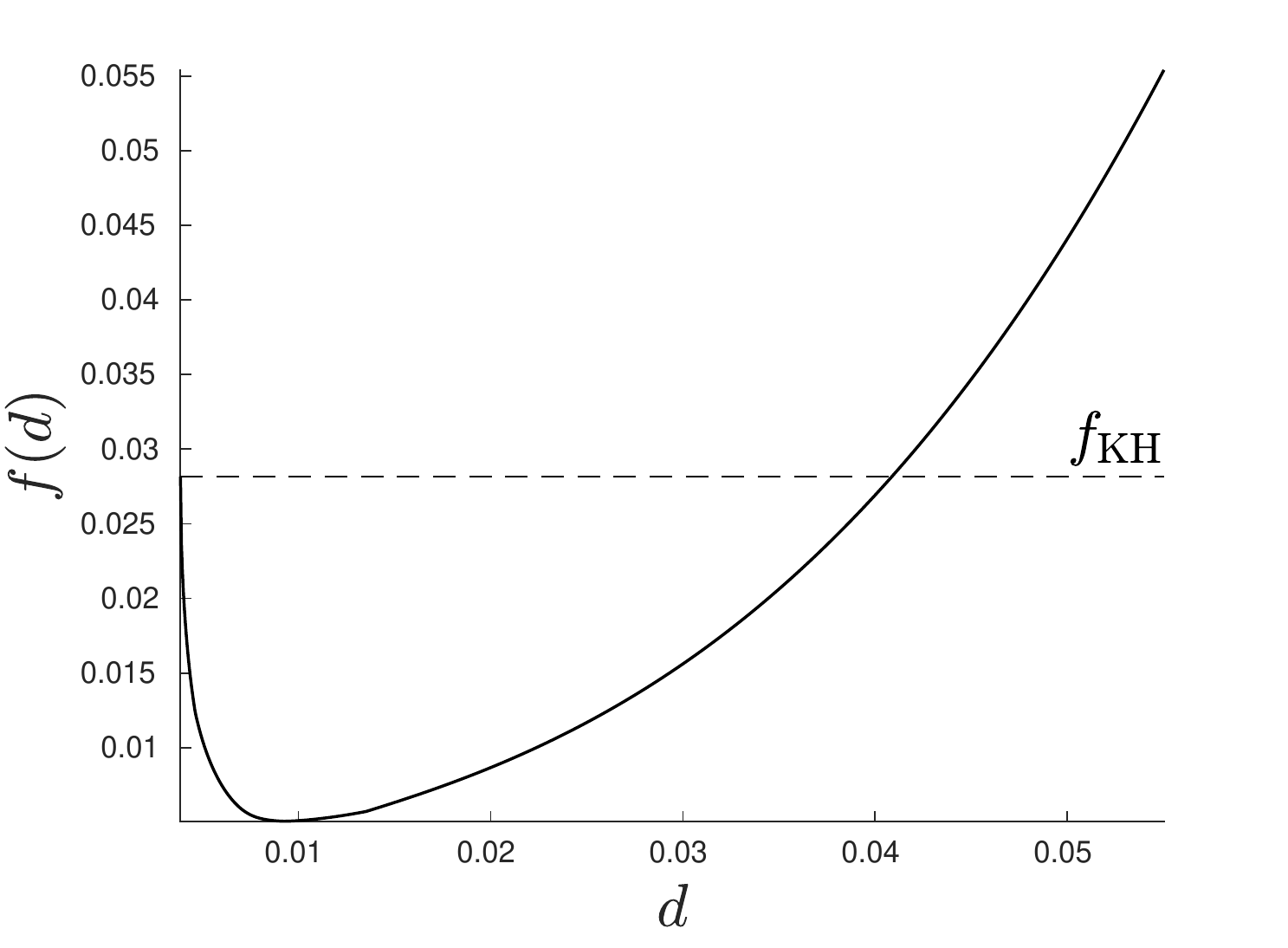}
	\caption{Trend of $f(\rsoivar)$ defined in \eqref{optnorm} in
		the neighbourhood of its minimum and comparison with $f_{\rm
			KH}$ defined in \eqref{norm_helio}. On the left
		$\delta=194\,\mbox{degrees}$, on the right
		$\delta=209\,\mbox{degrees}$; in both cases
		$\beta=105\,\mbox{degrees}$ and $C=2.97$.}
	\label{tfunex}
\end{figure}

In Fig. \ref{timeInRsoi} left we show the time spent inside the sphere
of influence. 
Here, we selected $C=2.97$, but the behaviour is similar for the other
values of the Jacobi constant.  Warmer colours denote longer encounter
times, and the dark-blue region corresponds to $\rsoi=0$. We can
clearly distinguish the bright-yellow region, corresponding to
encounters that last more than 100 hours. Outside this region the
duration of the encounter is considerably smaller. By comparison
with Fig. \ref{timeInRsoi} right, we can notice that the time spent
inside the sphere of influence is very short when the ratio
$\rsoi/\q3bp$ is close to $1$.  If we imposed a smaller value of
$\fPpcmmin$ in \eqref{trueanomthres}, the region in which
$\rsoi\sim\q3bp$ would be larger, replacing part of the dark-blue
region. In these cases, the patched-conic method gives only a slight
improvement with respect to a simple heliocentric Keplerian
propagation. This can be inferred from the example in
Fig. \ref{tfunex}. On the left, we show the graph of $f(\rsoivar)$ in
the neighbourhood of its minimum value for a point $(\beta,\delta)$
such that $\rsoi/\q3bp \sim 1$. Note that the minimum is very close to
$f_{\rm KH}$ and it belongs to a very small interval of values of
$\rsoivar$ where $f(\rsoivar)$ is not smooth. On the contrary, the
difference between the minumum of $f(\rsoivar)$ and $f_{\rm KH}$ is
significant for points $(\beta,\delta)$ in the region near the
collision curve, as shown in Fig. \ref{tfunex} right.  After
performing some tests, we chose $\fPpcmmin=5\,\mbox{degrees}$, but
higher values seem suitable too.

\begin{figure}
	\centering
	\includegraphics[width=.48\textwidth]{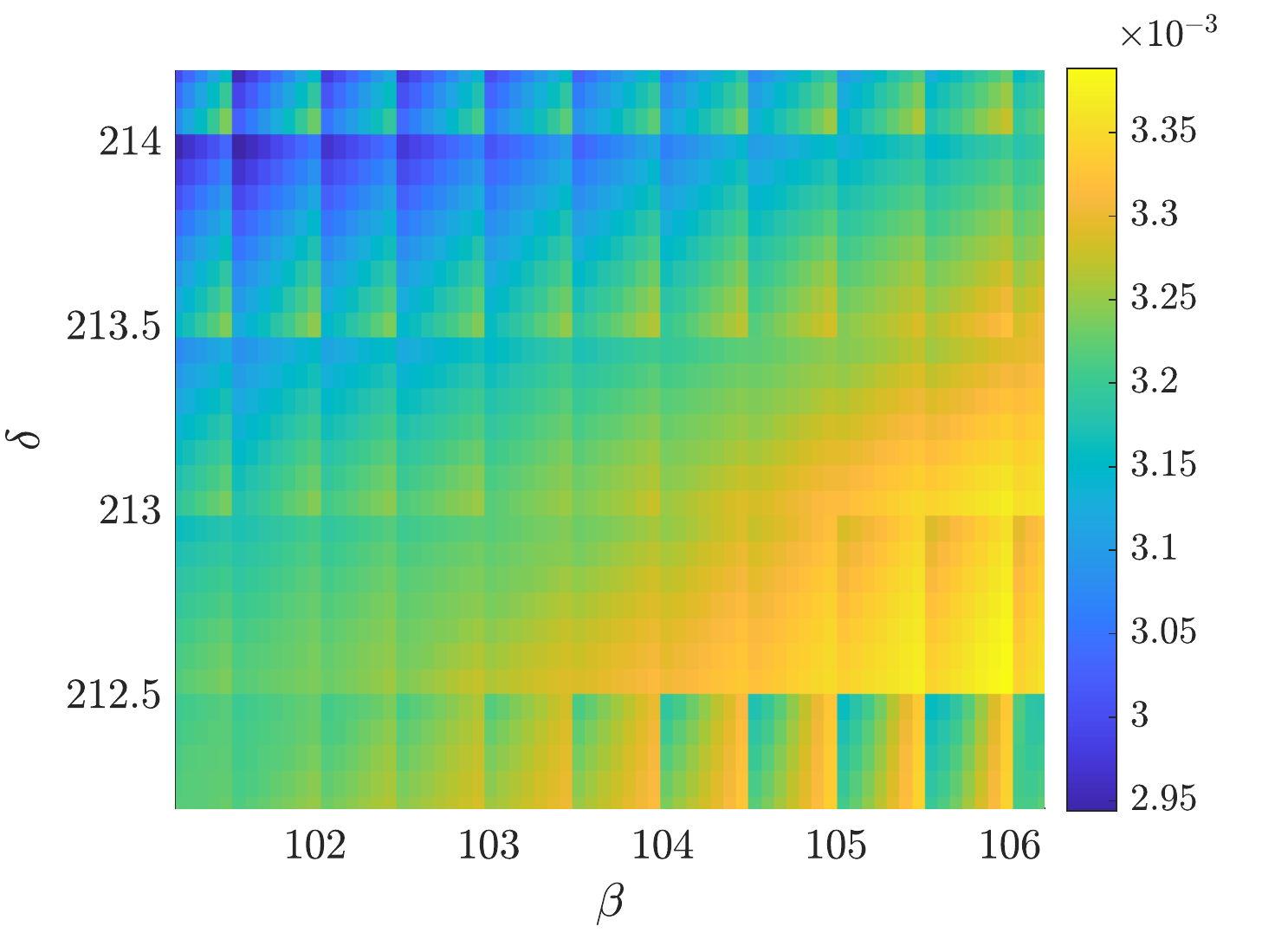}
	\caption{  Values of the target function \eqref{optnorm} computed for
		interpolated $\rsoi$ in a region in the $(\beta,\delta)$
		plane close to the collision curve for $C=2.963$.}
	\label{fig:interp}
\end{figure}

\begin{figure}
	\centering
	\includegraphics[width=.48\textwidth]{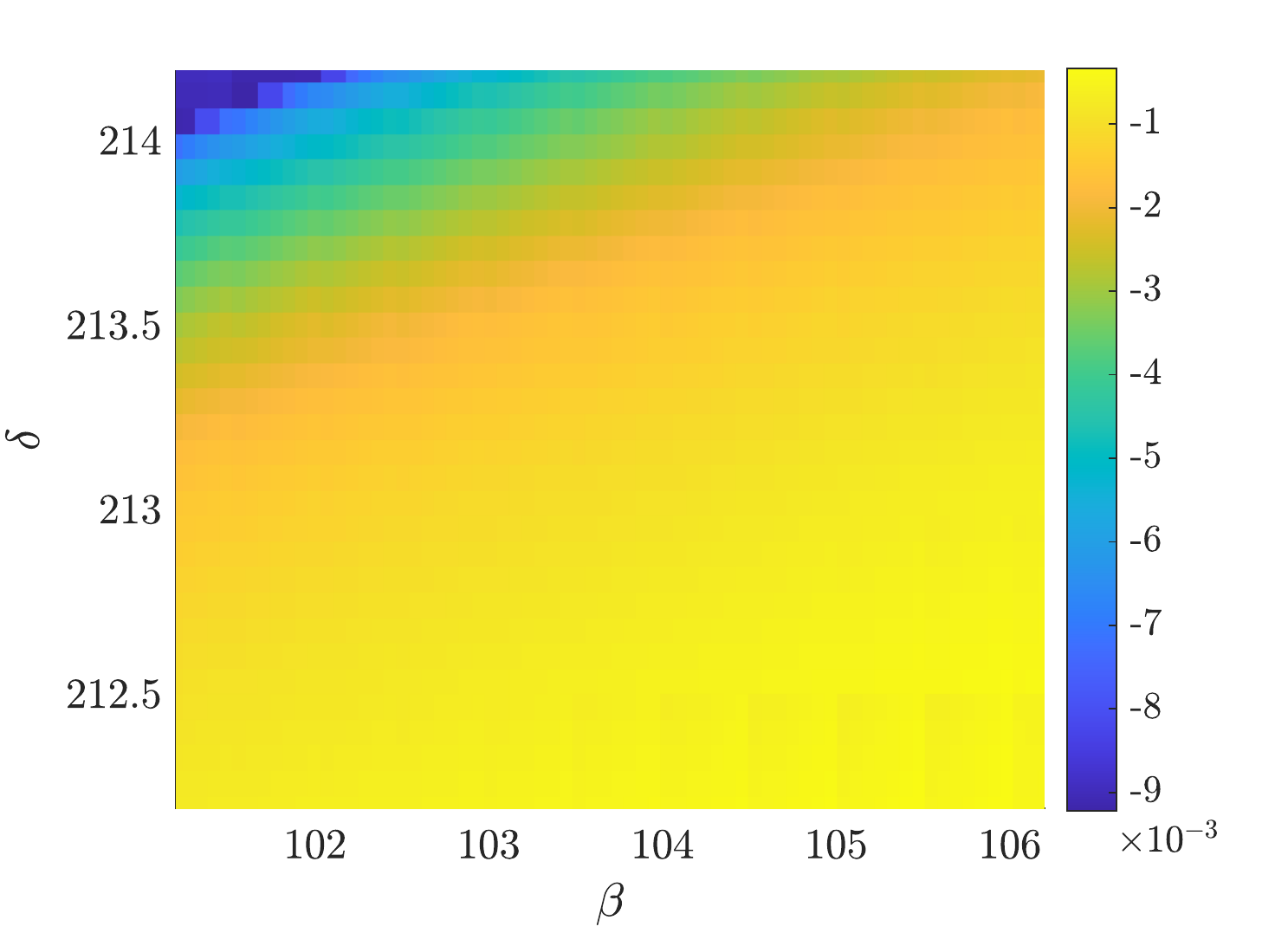}
	\includegraphics[width=.48\textwidth]{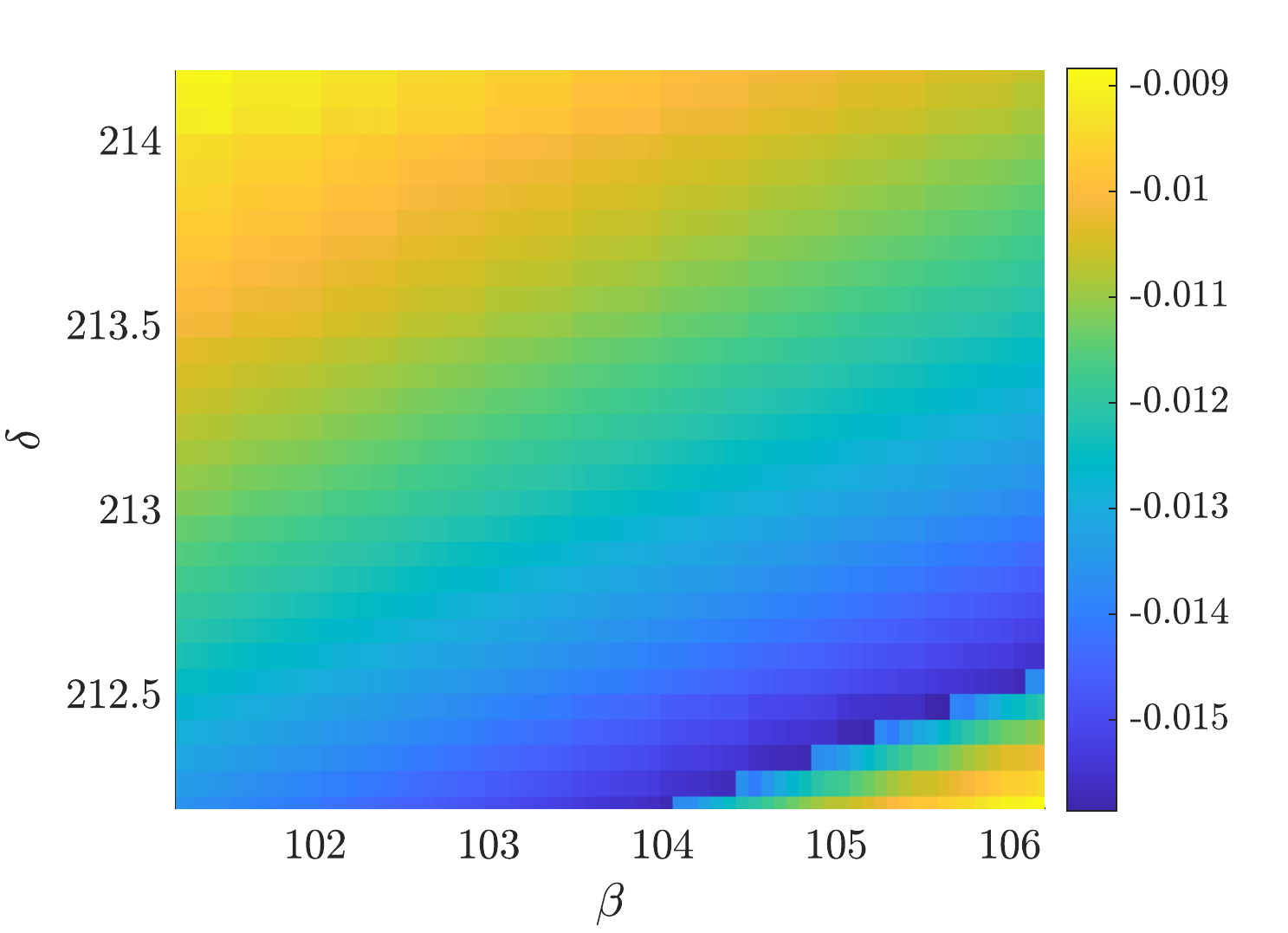}
	\caption{Comparison between the value of the target function
		\eqref{optnorm} computed with an interpolated $\rsoi$ and
		with Hill's (left) and Laplace's (right) radii. The considered region in the $(\beta,\delta)$ plane is the same as in Figure \eqref{fig:interp} and the value of $C$ is the same too.}
	\label{fig:compare}
\end{figure}

\begin{figure}
	\centering
	\includegraphics[width=.48\textwidth]{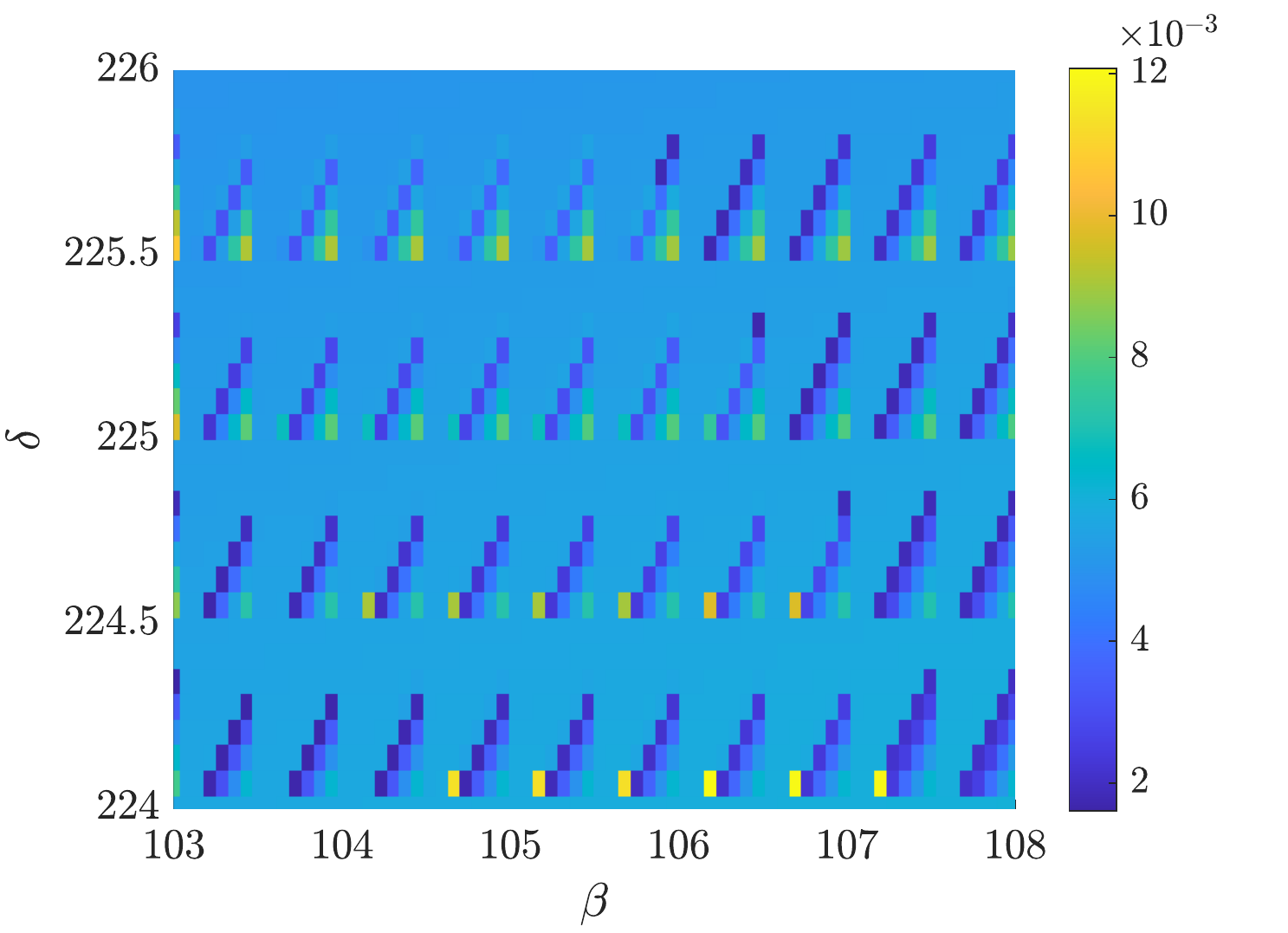}
	\caption{Values of the target function \eqref{optnorm} computed for
		interpolated $\rsoi$ in a region in the $(\beta,\delta)$
		plane far from the collision curve and $C=2.963$.}
	\label{fig:interp2}
\end{figure}

Using the interpolation technique explained in Section
\ref{sec:interp} we can compute $\rsoi$ for
values of $(C,\beta,\delta)$ different from the nodes of our
database. In Fig. \ref{fig:interp} we display the target
function \eqref{optnorm} computed with interpolated values of
$\rsoi$. In this example, we consider a region close to the collision
curve. Note that $f(\rsoi)$ remains confined to small values. 
In Fig. \ref{fig:compare} we compare
the radius $\rsoi$ obtained by interpolation with Hill's and Laplace's radii ($\rhill$,
$\rlapl$) by plotting the differences
\begin{equation*}
f(\rsoi) - f(\rhill), \qquad f(\rsoi) - f(\rlapl).
\end{equation*}
In this example we can see that both quantities above are always
negative. This means that, according to our selected norm, the
interpolated $\rsoi$ gives a better approximation than both Hill's and
Laplace's radii. In Fig. \ref{fig:interp2}, we repeat the same
experiment for $(\beta,\delta)$ points selected in a region farther
from the collision curve, where $\rsoi/\q3bp \sim 1$. The results
obtained are not as good as in the previous case. This is a
consequence of the features of $f(\rsoivar)$: as already shown, the
minimum of $f(\rsoivar)$ typically belongs to a very small
neighbourhood of values of $\rsoivar$ where the function is not
smooth. Thus, computing $\rsoi$ through an interpolation process can
produce a significant error. We could reduce the error by increasing the
density of the points in the database. However, this would also imply
increasing significantly the computational cost required for the
generation of the grid, which is not worthwhile since the minimum of
$f(\rsoi)$ is slightly smaller than $f_{\rm KH}$. This difficulty
related to the application of the interpolation process is an
additional reason for increasing the value of $\fPpcmmin$.

Our database is available at the webpage {\tt
	http://adams.dm.unipi.it/$\sim$cmg/rsoi/rsoi.html}.

\section{Conclusions}
\label{sec:conclude}
In this work we searched for an appropriate choice of the radius
$\rsoi$ of the sphere of influence of the Earth within a planar
patched-conic model. We developed an optimisation method, minimizing a
target function that depends on the possible radius of the sphere of
influence. This target function is made up of three components: 1) the
sup-norm of the difference between the 3-body and the patched-conic
orbits computed over a time interval including the encounter; 2) the
distance between the final states of the two orbits; 3) the norm of
the difference between the states at the minimum geocentric
distance. This procedure was repeated for a grid of different initial
conditions, defined by the Jacobi constant and by two angles, and
allowed us to construct a database of values of $\rsoi$. The resulting
data can be used to define a sphere of influence for generic initial
conditions, by interpolation.  We found that the best choice of
$\rsoi$ is typically close to Hill's radius $\rhill$, or larger. This
is consistent with the outcomes of \citet{Amato2017}. Some exceptions
can be found for initial conditions leading to a very deep close
encounter, or a collision.  Our procedure can be applied to the
3-dimensional case, as described in Appendix \ref{sec:3dim}. However,
the higher number of initial conditions that we should consider in
this case requires some care, both in the implementation and in the
data management. In future, it would also be interesting to
investigate the cases to which our method cannot be applied to more
deeply understand the phenomena causing the failure.

\section*{Acknowledgments}
This work was partially supported through the H2020 MSCA ETN
Stardust-Reloaded, Grant Agreement Number 813644.  CG, GFG and GB also
acknowledge the project MIUR-PRIN 20178CJA2B ``New frontiers of
Celestial Mechanics: theory and applications". The authors also
acknowledge the GNFM-INdAM (Gruppo Nazionale per la Fisica
Matematica).

\appendix

\section{Classical definitions of the sphere of influence}
\label{app:rsoi}

The classical definitions of the sphere of influence are 
\begin{enumerate}
	\item Laplace's sphere, given in equation \eqref{rlaplace}, approximating the radius of a spherical region where the
	perturbing effect of the Sun on the particle's planetocentric
	orbit is lower than the perturbing effect of the planet on the
	particle's heliocentric orbit \citep{Battin_1999}; 
	\item Hill's sphere, defined in equation \eqref{rhill}, approximating the distance from the planet to the Lagrangian point
	$L_1$ (\citet{Chebotarev1964}).
\end{enumerate}
In \citet{Chebotarev1964}, the author introduced a further definition
of sphere of influence, corresponding to "the region of space within
which the attraction of the planet is greater than solar attraction",
see also \citet{Souami2020}.  The radius of Chebotarev's sphere is
\begin{equation*}
\rcheb=\rho\Big(\frac{m_{2}}{m_{1}}\Big)^{1/2}.
\end{equation*}

As shown in \citet{Souami2020}, relation
\begin{equation}
\rcheb<\rlapl<\rhill
\label{ineq_rsoi}
\end{equation}
holds when $m_2\ll m_1$. In particular, $\rhill>\rlapl$ if $m_2/m_1<1/243$
and $\rlapl>\rcheb$ if $m_2/m_1<1$. Since the ratio between the mass of
Jupiter and that of the Sun is about 0.001, \eqref{ineq_rsoi} is true
for all solar system planets.

\subsection{Laplace's sphere}
\label{app:laplace}
We derive the radius of Laplace's sphere of action following
\citet{Battin_1999}. We consider the Sun-Earth problem, but what
follows is valid for any planet. We adopt the same non-dimensional
system described in Section \ref{sec:CR3BP}, and we denote by
${\boldsymbol\rho}$ the heliocentric position vector of the Earth and
by ${\bm r}$ and ${\bm d}$ the position of the small body
with respect to the Sun and the Earth, respectively. Let
\begin{equation*}
d=|{\bm  d}|,\qquad r=|{\bm  r}|,\qquad \rho=|{\boldsymbol\rho}|=1.
\end{equation*}
The small body is approximated as a massless particle. 
We can consider two perturbed two-body problems: the Sun-particle problem perturbed by the Earth and the Earth-particle problem perturbed by the Sun. These are described by
\begin{align}
\frac{d^2\bm r}{dt^2}& = -(1-\mu)\frac{\bm 
	r}{r^3}-\mu\left(\frac{\bm 
	d}{d^3}+\frac{\boldsymbol{\rho}}{\rho^3}\right) = {\bm 
	a}_{S,k}+{\bm  a}_{S,p},
\label{acc1}\\
\frac{d^2\bm  d}{dt^2}& = -\mu\frac{\bm  d}{d^3}-(1-\mu)\left(\frac{\bf r}{r^3}-\frac{\boldsymbol{\rho}}{\rho^3}\right) = {\bm  a}_{E,k}+{\bm  a}_{E,p},
\label{acc2}
\end{align}
where ${\bm  a}_{S,k}$, ${\bm  a}_{E,k}$ are the two-body accelerations
and ${\bm  a}_{S,p}$, ${\bm  a}_{E,p}$ are the perturbing accelerations. 

Laplace's sphere of influence is obtained by finding the geocentric distance
where the ratios between two-body and perturbing
accelerations in \eqref{acc1} and \eqref{acc2} are the same, that is
\begin{equation*}
\frac{a_{S,p}}{ a_{S,k}} = \frac{a_{E,p}}{a_{E,k}}.
\end{equation*}
This condition becomes
\begin{equation*}
\frac{\mu}{1-\mu}\frac{r^2}{d^2}\sqrt{1+d^4+2d^2\cos\varphi}
=
\frac{1-\mu}{\mu}\frac{d^2}{r^2}\sqrt{1+r^4-2r^2\cos\alpha}.
\end{equation*}
Noting that
\begin{equation*}
r\cos\alpha + d\cos\varphi = 1,
\end{equation*}
we can write
\begin{equation*}
\cos\alpha = \frac{1}{r}-\frac{d}{r}\cos\varphi,
\end{equation*}
so that 
\begin{equation*}
\frac{d^4}{r^4} =
\frac{\mu^2}{(1-\mu)^2}\sqrt{\frac{1+d^4+2d^2\cos\varphi}{1+r^4- 2r\left(1-d\cos\varphi\right)}}.
\end{equation*}
Since $\mu\ll1$, the sphere of influence will
be much closer to the Earth than to the Sun, so that we can assume
$d\ll 1$ and therefore expand in power series of $d$.
Noting also that
\begin{equation}
r^2 = 1+d^2-2d\cos\varphi,
\label{rrhod}
\end{equation}
we have
\begin{equation*}
\begin{aligned}
\sqrt{1+r^4-2r\left(1-d\cos\varphi\right)} &=
d\sqrt{1+3\cos^2\varphi-4d\cos\varphi + \mathcal
	O\left(d^2\right)},\\ \sqrt{1+d^4+2d^2\cos\varphi}
&= 1+\mathcal O\bigl(d^2\bigr)
\end{aligned}
\end{equation*}
and
\begin{equation*}
r^4 = \left(1+d^2-2d\cos\varphi\right)^2 = 1+\mathcal O\bigl(d^2\bigr).
\end{equation*}
Then, we can write
\begin{equation*}
\begin{aligned}
d^5 &= \frac{\mu^2}{(1-\mu)^2}\frac{1}{(1+3\cos^2\varphi-4d\cos\varphi)^{1/2}}+\mathcal O\left(d\right)\\
&=\frac{\mu^2}{(1-\mu)^2}\frac{1}{(1+3\cos^2\varphi)^{1/2}}+ \mathcal O\left(d\right),
\end{aligned}
\end{equation*}
and
\begin{equation*}
d = \left(\frac{\mu^2}{(1-\mu)^2}\right)^{1/5}\frac{1}{(1+3\cos^2\varphi)^{1/10}}.
\end{equation*}
Considering that $1\le (1+3\cos^2\varphi)^{1/10} \le 2^{2/5}$, we simply
approximate it with $1$ and finally get
the value of the radius of Laplace's sphere of influence
\begin{equation*}
d = \left(\frac{\mu}{1-\mu}\right)^{2/5}\sim\mu^{2/5}.
\end{equation*}
The approximation performed is typical in modern works, see for example \citet{Battin_1999}; on the contrary, \citet{Laplace1805} chose the alternative $(1+3\cos^2\varphi)^{1/10} \sim 2^{1/5}$, leading to
\begin{equation*}
d = \left(\frac{\mu}{\sqrt{2}(1-\mu)}\right)^{2/5}.
\end{equation*}
\subsection{Hill's sphere}
\label{app:hill}
In the planar CR3BP, the region of allowed motion is given
by the Hill region. Depending on its energy, the small body can be
free to move everywhere, or be confined to smaller regions. In
particular, if its energy is lower than the energy
associated to the Lagrangian point $L_1$, then the region of possible
motion is closed and a small body in the vicinity of the planet is
bound to stay close to the planet.

The definition of the radius of the Hill sphere of influence comes
from an approximation of the distance of the Lagrangian point $L_1$
from the planet.

Equilibrium points are found by solving the system
\begin{equation*}
\begin{cases}
\dot x = p_x+y = 0\\ \dot y = p_y-x = 0\\ \dot p_x =
p_y-\frac{(1-\mu)(x+\mu)}{[(x+\mu)^2+y^2]^{(3/2)}}-\frac{\mu(x-1+\mu)}{[(x-1+\mu)^2+y^2]^{(3/2)}}
= 0\\ \dot p_y =
-p_x-\frac{(1-\mu)y}{[(x+\mu)^2+y^2]^{(3/2)}}-\frac{\mu
	y}{[(x-1+\mu)^2+y^2]^{(3/2)}} = 0
\end{cases} .
\end{equation*}
From the first two equations we get $p_x=-y, p_y=x$. Setting $y=0$ to
search for collinear Lagrangian points, from the third equation we get
\begin{equation*}
x=\frac{1-\mu}{\lvert x+\mu\rvert(x+\mu)}+\frac{\mu(x-1+\mu)}{\lvert
	x-1+\mu\rvert(x-1+\mu)}.
\end{equation*}
Let $\tilde x=x-1+\mu$. We notice that $-1<\tilde x<0$ for the
Lagrangian point $L_1$ located between the primary and the secondary
body. The equation above yields an explicit expression for $\mu$. For
$L_1$ this is
\begin{equation*}
\mu=\frac{\tilde x^5+3\tilde x^4+3\tilde x^3}{\tilde x^4+2\tilde
	x^3+\tilde x^2+2\tilde x+1},
\end{equation*}
which, after expanding around $\tilde x=0$, becomes
\begin{equation*}
\mu=3\tilde x^3-3\tilde x^4+4\tilde x^5+\mathcal O(\tilde x^6).
\end{equation*}
The distance $\tilde x$ of the $L_1$ point from the planet is obtained
by solving the equation above in power series of $\mu^{1/3}$
\begin{equation*}
\tilde
x=\left(\frac{\mu}{3}\right)^{1/3}-\frac{1}{3}\left(\frac{\mu}{3}\right)^{2/3}-\frac{\mu}{27}+\mathcal
O(\mu^{4/3}),
\end{equation*}
see \citep{Szebehely_1967}.
The radius of Hill's sphere corresponds to the first
term of this expansion.

\subsection{Chebotarev's sphere}
\label{app:chebotarev}

Chebotarev's sphere is obtained by imposing the condition 
\begin{equation}
a_{S,k}=a_{E,k},
\label{chabcondition}
\end{equation}
with $a_{S,k}$ and $a_{E,k}$ defined in Section
\ref{app:laplace}. As a consequence of \eqref{rrhod}, where the
Sun-Earth distance has been normalized to 1, it holds
\begin{equation*}
r^ 2 = 1+\mathcal{O}\left(d\right)
\end{equation*}
so that
\begin{equation*}
a_{S,k} = \frac{1-\mu}{r^2}.
\end{equation*}
Since
\begin{equation*}
a_{E,k} = \frac{\mu}{d^2}
\end{equation*}
from condition \eqref{chabcondition}, we obtain
\begin{equation*}
d = \Big(\frac{\mu}{1-\mu}\Big)^{1/2}. 
\end{equation*}

\section{Jacobi integral}
\label{app:jacobi}
The expressions of the Jacobi integral given in 
\eqref{Cdef}, \eqref{Chelio} and \eqref{Cplaneto} are derived below. 

\subsection{Barycentric inertial coordinates}
As a first step, it is convenient to consider the coordinate change
$(x,y,z)\mapsto(X,Y,Z)$ defined by
\begin{equation}
\left( \begin{array}{c} X \\ Y \\ Z\end{array} \right) =
\left[ \begin{array}{ccc} \cos\lambda & -\sin\lambda & 0
\\ \sin\lambda & \cos\lambda & 0 \\ 0 & 0 & 1 \end{array}
\right]\left( \begin{array}{c} x \\ y \\ z\end{array} \right),
\label{syn2bar}
\end{equation}
where $\lambda$ is the longitude of the Earth, given by 
\begin{equation*}
\lambda = \lambda_{0} + t - t_0,
\end{equation*}
with $t$ the non-dimensional time, $t_0$ its initial value and
$\lambda_0$ the value of $\lambda$ at $t_0$. Note that if $t_0=0$ and
$\lambda_0=0$ the value of the longitude corresponds to the value of
the non-dimensional time. The coordinates $X,Y,Z$ are
barycentric inertial coordinates.  Differentiating \eqref{syn2bar}
with respect to $t$ gives the transformation of the velocity
components:
\begin{equation}
\left( \begin{array}{c} \dot{X} \\ \dot{Y} \\ \dot{Z}\end{array}
\right) = \left[ \begin{array}{ccc} \cos\lambda & -\sin\lambda & 0
\\ \sin\lambda & \cos\lambda & 0 \\ 0 & 0 & 1 \end{array}
\right]\left( \begin{array}{c} \dot{x}-y \\ \dot{y}+x
\\ \dot{z}\end{array} \right).
\label{syn2bardot}
\end{equation}
From \eqref{Cdef}, applying the transformations \eqref{syn2bar} and \eqref{syn2bardot}, we obtain
the Jacobi integral as a function of the barycentric inertial
variables:
\begin{equation*}
\jacobi = 2(X\dot{Y}-Y\dot{X}) +
\frac{2(1-\mu)}{\bar{r}}+\frac{2\mu}{\bar{d}}-V^2, \label{Cbar}
\end{equation*}
with
\begin{equation*}
\begin{split}
\bar{r} & = 
\sqrt{(X+\mu\cos\lambda)^2+(Y+\mu\sin\lambda)^2+Z^2},\\ \bar{d}
& = 
\sqrt{(X-(1-\mu)\cos\lambda)^2+(Y-(1-\mu)\sin\lambda)^2+Z^2},\\ V
& =  \dot{X}^2+\dot{Y}^2+\dot{Z}^2.
\end{split}
\end{equation*}
It is straightforward in the new variables to express the heliocentric
position and velocity vectors $\bm{R}$ and
$\dot{\bm{R}}$ of the small body, as well as the planetocentric ones,
$\bm{D}$ and $\dot{\bm{D}}$. It holds
\begin{align}
\bm{R} & =  (X+\mu\cos\lambda, Y+\mu\sin\lambda, Z)^T,
\label{helioposvec}\\
\dot{\bm{R}} & =  (\dot{X}-\mu\sin\lambda,  \dot{Y}+\mu\cos\lambda,\dot{Z})^T, 
\label{heliovelvec}\\
\bm{D} & =  (X-(1-\mu)\cos\lambda, Y-(1-\mu)\sin\lambda, Z)^T,
\label{planetoposvec}\\
\dot{\bm{D}} &  =  (\dot{X}+(1-\mu)\sin\lambda, \dot{Y}-(1-\mu)\cos\lambda,\dot{Z})^T.
\label{planetovec}
\end{align}

\subsection{Heliocentric elements} 
Since
\begin{equation*}
{\dot{\bm{R}}}\cdot{\dot{\bm{R}}} =
(1-\mu)\left(\frac{2}{\bar{r}}-\frac{1}{\smaH}\right)
\end{equation*}
and
\begin{equation*}
\begin{split}
{\dot{\bm{R}}}\cdot{\dot{\bm{R}}} =
V^2+2\mu(-\dot{X}\sin\lambda+\dot{Y}\cos\lambda)+\mu^2,	\end{split}
\end{equation*}
from \eqref{heliovelvec}, we have
\begin{equation*}
V^2 = \frac{2(1-\mu)}{\bar{r}}-\frac{1-\mu}{\smaH} -\mu^2
-2\mu(-\dot{X}\sin\lambda+\dot{Y}\cos\lambda).
\label{subs1_jachelio}
\end{equation*}
Furthermore, it holds
\begin{equation*}
(\bm{R}\times\dot{\bm{R}})\cdot \bm{e}_Z = \sqrt{(1-\mu)\smaH(1-\eccH^2)}\cos\inclH,
\end{equation*}
with $\bm{e}_Z = (0,0,1)^T$. From
\eqref{helioposvec} and \eqref{heliovelvec}, we get
\begin{equation*}
\begin{split}
(\bm{R}\times\dot{\bm{R}})\cdot \bm{e}_Z &=
X\dot{Y}-Y\dot{X}+\mu(-\dot{X}\sin\lambda +
\dot{Y}\cos\lambda)\\ & \quad +\mu(X\cos\lambda+Y\sin\lambda)+\mu^2,
\end{split}
\end{equation*}
so that 
\begin{equation}
\begin{split}
X\dot{Y}-Y\dot{X} &= \sqrt{(1-\mu)\smaH(1-\eccH^2)}\cos \inclH
-\mu(-\dot{X}\sin\lambda+\dot{Y}\cos\lambda)\\
&\quad -\mu(X\cos\lambda+Y\sin\lambda)-\mu^2.
\end{split}
\label{subs2_jachelio}
\end{equation}
Substituting \eqref{subs1_jachelio} and \eqref{subs2_jachelio} in
\eqref{Cbar} and simplifying yields
\begin{equation}
\begin{split}
\jacobi & =   \frac{1-\mu}{\smaH} +2\sqrt{(1-\mu)\smaH(1-\eccH^2)}\cos
\inclH+\frac{2\mu}{\bar{d}}\\
&\quad  +2\mu(X\cos\lambda+Y\sin\lambda) -\mu^2.
\end{split}
\label{jacnew}
\end{equation}
Equation \eqref{Chelio} is obtained using $\bar{d}=d$ and 
\begin{equation*}
X\cos\lambda+Y\sin\lambda = x = r\cos\alpha -\mu,
\end{equation*}
in \eqref{jacnew}, with $\alpha$ the angle between the small body and the Earth seen from
the Sun.

\subsection{Planetocentric elements} 
Let us assume that the osculating planetocentric elements correspond
to the orbital elements of hyperbolic trajectories, i.e. they are such
that $\eccG>1$. Then, we have
\begin{equation*}
\dot{\bm{D}}\cdot \dot{\bm{D}} =
\mu\left(\frac{2}{\bar{d}}+\frac{1}{\smaG}\right)
\end{equation*}
and
\begin{equation*}
(\bm{D}\times\dot{\bm{D}})\cdot \bm{e}_Z =
\sqrt{(1-\mu)\smaG(\eccG^2-1)}\cos \inclG.
\end{equation*}
Since from \eqref{planetoposvec} and
\eqref{planetovec} 
\begin{equation*}
\begin{split}
\dot{\bm{D}}\cdot\dot{\bm{D}} =
V^2+2(1-\mu)(\dot{X}\sin\lambda-\dot{Y}\cos\lambda)+(1-\mu)^2
\end{split}
\end{equation*}
and
\begin{equation*}
\begin{split}
(\bm{D}\times\dot{\bm{D}})\cdot \bm{e}_Z &=
X\dot{Y}-Y\dot{X}+(1-\mu)(\dot{X}\sin\lambda-
\dot{Y}\cos\lambda)\\ &\quad -(1-\mu)(X\cos\lambda+Y\sin\lambda)+(1-\mu)^2,
\end{split}
\end{equation*}
we obtain
\begin{equation}
V^2 = \frac{2\mu}{\bar{d}}+\frac{\mu}{\smaG} -(1-\mu)^2
-2(1-\mu)(\dot{X}\sin\lambda-\dot{Y}\cos\lambda)
\label{subs1_jacplaneto}
\end{equation}
and 
\begin{equation}
\begin{split}
X\dot{Y}-Y\dot{X} & =  \sqrt{(1-\mu)\smaG(\eccG^2-1)}\cos\inclG-(1-\mu)(\dot{X}\sin\lambda-\dot{Y}\cos\lambda)\\ & \quad 
+(1-\mu)(X\cos\lambda+Y\sin\lambda)-(1-\mu)^2.
\end{split}
\label{subs2_jacplaneto}
\end{equation}
Substituting \eqref{subs1_jacplaneto} and \eqref{subs2_jacplaneto} in
\eqref{Cbar} and simplifying, we find
\begin{equation}
\begin{split}
\jacobi & =  -\frac{\mu}{\smaG} +2 \sqrt{(1-\mu)\smaG(\eccG^2-1)}\cos \inclG
+\frac{2(1-\mu)}{\bar{r}}\\
& \quad +2(1-\mu)(X\cos\lambda+Y\sin\lambda)
-(1-\mu)^2.
\end{split}
\label{jacnew2}
\end{equation}
Equation \eqref{Cplaneto} is obtained using $\bar{r}=r$ and
\begin{equation*}
X\cos\lambda+Y\sin\lambda = x = -d\cos\varphi+ 1 -\mu,
\end{equation*}
in \eqref{jacnew2}, with $\varphi$ the angle between the small body and the Sun seen from
the Earth.

\section{Collisions}
\label{app:coll}
Let us fix a value $C$ of the Jacobi integral and an initial distance
$d_0$ from the Earth, as in Section~\ref{sec:ic_planar}. We describe
the procedure applied to compute the collision curve displayed in
Fig. \ref{fig:rsoi}, defined by the values of the angles $(\beta,
\delta)$ such that the corresponding 3-body trajectory passes through
the centre of the Earth.  In the regularised variables, this condition
corresponds to
\begin{equation*}
u(\tau;\beta,\delta) = 0,\qquad v(\tau;\beta,\delta) = 0 
\end{equation*}
for some value $\tau>0$ of the fictitious time. Let
\begin{equation*}
\bm{F}(\beta,\delta,\tau) =
\left(
\begin{array}{c}
u(\tau;\beta,\delta)\\
v(\tau;\beta,\delta)\\
\end{array}
\right)
\end{equation*}
and assume that $(\beta_c,\delta_c,\tau_c)$ is a collision condition, i.e. $\bm{F}(\beta_c,\delta_c,\tau_c)= 0$. From the implicit
function theorem, if the Jacobian matrix
\begin{equation}
D\bm{F} =
\frac{\partial\bm{F}}{\partial(\beta,\delta)} 
\label{eqn:jacMat}
\end{equation}
is non-singular in $(\beta_c,\delta_c,\tau_c)$, then there exists a
local parametrization 
\[
\tau \mapsto {\bm w}(\tau) :=\left(\begin{array}{c}
w_1(\tau)\\w_2(\tau)
\end{array}\right)
\]
of the
collision curve defined in a neighbourhood $I$ of $\tau_c$, such that 
\begin{equation*}
\begin{split}
&w_1(\tau_c)= \beta_c,\qquad w_2(\tau_c)= \delta_c,\\
& \bm{F}(\bm{w}(\tau),\tau)=0, \qquad \forall \tau\in I.
\end{split}
\end{equation*}
Moreover, the tangent vectors to the collision curve can be written as 
\begin{equation*}
\bm{w}'(\tau) =
[D\bm{F}]^{-1}(w_1(\tau),w_2(\tau),\tau)
\frac{\partial\bm{F}}{\partial\tau}
(w_1(\tau),w_2(\tau),\tau).
\end{equation*}
In our case the Jacobian matrix \eqref{eqn:jacMat} can be computed as
\begin{equation*}
D\bm{F}= \left(\frac{\partial(p_u,p_v,u,v)}{\partial(\beta,\delta)}\,\bm{\mathsf{M}} \,{\boldsymbol{\Psi}}\right)^{\rm T},
\end{equation*}
where
\begin{equation*}
\bm{\mathsf{M}}= \left[\begin{array}{c c c c}
1 & 0 & 0 &  0\\
0 & 0 & 1 & 0
\end{array}\right],
\end{equation*}
$\boldsymbol{\Psi}$ is the state transition matrix of the system with
Hamiltonian $K$ (see \ref{eqn:LCham}), and $\rm T$ stands for vector transposition. 
Moreover,
\[
\frac{\partial\bm{F}}{\partial\tau} =
\frac{\partial K}{\partial (p_u,p_v)}.
\]
To compute the collision curve, we perform the following steps: 
(i) we select a starting point $(\beta_c,\delta_c)$ that
leads to a collision at time $\tau_c$; (ii) we move slightly away from $(\beta_c,\delta_c)$ along
the direction of the tangent vector ${\bm w}(\tau_c)$ and
look for a new collision point in the direction orthogonal
to ${\bm w} (\tau_c)$; (iii) we iterate the
process once the new collision point has been found.

\section{3-dimensional case}
\label{sec:3dim}

To apply the method described in Section \ref{sec:newdef} to the 3-dimensional problem, a few
changes are required. In particular, it is necessary to extend the set of
initial conditions and use the KS regularisation for the 3-body
propagation. Moreover, since the initial conditions will be
described by more than three variables, a multi-variate interpolation
method is necessary. More details on the selection of the initial
conditions and a brief summary of the KS regularisation are reported
below.

\subsection{Initial Conditions}

In the 3-dimensional configuration space, the initial position of
the particle must be selected on the sphere $\mathcal{S}_0$ defined as
\begin{equation*}
\mathcal{S}_0 =\{(x,y,z)\in \mathbb{R}: (x-1+\mu)^2+y^2+z^2=d_0^2\}.
\end{equation*}
As in the planar case, for any initial position $(x_0,y_0,z_0)$ the
norm $v_0$ of all possible velocities is completely
determined once the Jacobi constant $C$ has been fixed:
\begin{equation}
v_0^2 = \dot{x}_0^2+\dot{y}_0^2+\dot{z}_0^2 =
2\frac{1-\mu}{r_0}+2\frac{\mu}{d_0}+x_0^2+y_0^2-C,
\label{v0def_3d}
\end{equation}
where $r_0=r(x_0,y_0,z_0)$ (see \ref{rdef}). In the 3-dimensional case, especially for higher values of $C$, we may have points
$(x_0,y_0)\in \mathcal{S}_0$ for which relation \eqref{v0def_3d} gives $v_0^2<0$: these points must be
excluded from the domain.

Four angular coordinates $\beta$, $\psi$, $\delta$ and $\xi$ are
required to parametrize the initial conditions.
The angles $\beta\in[0,2\pi)$ and $\psi\in[-\frac{\pi}{2},\frac{\pi}{2}]$ are the spherical coordinates with respect to the geocentric synodic reference frame and are used to define the position vector. 
Consider now a reference
frame $A\bar{x}\bar{y}\bar{z}$ centred at the position of the
small body $A$,
obtained through the transformation
\begin{equation*}
\left(\begin{array}{c}
\bar{x}+d_0\\\bar{y}\\\bar{z}\end{array}\right)=\left[\begin{array}{ccc}
\cos\beta\cos\psi & \sin\beta\cos\psi & \sin\psi\\
-\sin\beta & \cos\beta & 0\\
-\cos\beta\sin\psi & -\sin\beta\sin\psi& \cos\psi 
\end{array}\right]\left(\begin{array}{c}
x_P\\y_P\\z_P\end{array}\right).
\end{equation*}
The angles $\delta\in[0,2\pi)$ and $\xi\in[-\frac{\pi}{2},\frac{\pi}{2}]$ are spherical coordinates with respect to the reference frame $A\bar{x}\bar{y}\bar{z}$ and are used to define the velocity vector. 
Thus, the initial conditions
in the barycentric synodic variables are given by
\begin{align*}
x_0 &= 1-\mu + d\cos\beta\cos\psi, \\ y_0 &=
d\sin\beta\cos\psi, \\ z_0 &= 
d\sin\psi, \\ 
\dot{x}_0& = 
v_0(\cos\beta\cos\psi\cos\delta\cos\xi-
\sin\beta\sin\delta\cos\xi-\cos\beta\sin\psi\sin\xi),\\
\dot{y}_0& = 
v_0(\sin\beta\cos\psi\cos\delta\cos\xi +
\cos\beta\sin\delta\cos\xi-\sin\beta\sin\psi\sin\xi),\\
\dot{z}_0& =  v_0(\sin\psi\cos\delta\cos\xi +
\cos\psi\sin\xi).
\end{align*}
Note that if $\psi=0$ and $\xi=0$, we obtain the initial conditions of
the planar problem. Also for the 3-dimensional problem, we need to
consider $\delta \in \big(\frac{\pi}{2},\frac{3\pi}{2}\big)$ to obtain
trajectories entering the sphere $\mathcal{S}_0$. The set
of initial conditions is determined constructing a regularly-spaced
grid in the 4-dimensional space $(\beta,\psi,\delta,\xi)$.

\subsection{Propagation with KS variables}
\label{app:KS}
For the 3-dimensional problem, we need to perform the propagation
using the variables $(U_1,U_2,U_3,U_4,u_1,u_2,u_3,u_4)$ of the Kustaanheimo-Stiefel
regularisation \citep{Stiefel_1971}. We introduce these variables
through the transformation
\begin{equation*}
\left(\begin{array}{c} x-1+\mu\\y\\z\\0
\end{array}\right)=L(\KSu)\KSu, \qquad \left(\begin{array}{c}
\dot{x}\\\dot{y}\\\dot{z}\\0
\end{array}\right)=2L\frac{({\KSu)}}{\vert \KSu\vert^2}\KSUP,
\end{equation*}
where $\KSu=(u_1,u_2,u_3,u_4)^{\rm T}$, $\KSUP=(U_1,U_2,U_3,U_4)^{\rm T}$ and
\begin{equation*}
L(\KSu) = \left[\begin{array}{cccc} u_1 & -u_2 & -u_3 & u_4 \\ u_2 &
u_1 & -u_4 & -u_3 \\ u_3 & u_4 & u_1 & u_2 \\ u_4 & -u_3 & u_2 &
-u_1\end{array}\right].
\end{equation*}
We also introduce the fictitious time $\tau$ through the differential relation
\begin{equation*}
\frac{d\tau}{dt}=\frac{1}{\vert \KSu\vert^2}.
\end{equation*}
Let us denote with a prime the derivative with respect to $\tau$. For
the fixed value of the Jacobi integral $J=C$, the equations of motion
are
\begin{equation*}
\begin{split}
\KSu' & =  \KSUP,\\ \KSUP ' & =
-\frac{1}{4}\big(C-h(\KSu)\big)\KSu+L(\KSu)^{\rm T} B L(\KSu)\KSUP +
\frac{\vert\KSu\vert^2}{2}L(\KSu)^{\rm T} \bm{f}(\KSu),
\end{split}
\end{equation*}
where
\begin{align*}
& B = \left[\begin{array}{cccc} 0 & 2 & 0 & 0\\ -2 & 0 & 0 & 0 \\0 & 0 &
0 & 0 \\0 & 0 & 0 & 0
\end{array}\right], \\
& \bm{f}(\KSu)=\bm{\mathsf{K}} bL(\KSu)\KSu-\frac{1-\mu}{r^3}L(\KSu)\KSu, \\ &
h(\KSu)= 2\frac{1-\mu}{r}+\bm{\mathsf{K}}L(\KSu)\cdot \bm{\mathsf{K}}L(\KSu),\\
& r =  \sqrt{\vert \KSu\vert^2+1+2 \bm{e}_{u} \cdot L(\KSu)\KSu},\\
& \bm{e}_{u}=  (1, 0, 0, 0)^{\rm T}, \\
&  \bm{\mathsf{K}} = \left[\begin{array}{cccc} 1 & 0 &
0 & 0\\ 0 & 1 & 0 & 0\\0 & 0 & 0 & 0\\ 0 & 0 & 0 & 0
\end{array}\right].
\end{align*}
As for the planar problem, after the propagation we transform the
orbit back to barycentric synodic coordinates.

\newpage
\bibliographystyle{abbrvnat}
\bibliography{refs}

\end{document}